\documentclass[%
 reprint,
superscriptaddress,
showpacs,preprintnumbers,
 amsmath,amssymb,
 aps,
prb,
floatfix,
]{revtex4-1}
\usepackage{amsmath,amssymb}
\usepackage{graphicx}
\usepackage{dcolumn}
\usepackage{bm}

\usepackage{hyperref}
\hypersetup{
    unicode=false,          
    pdftoolbar=true,        
    pdfmenubar=true,        
    pdffitwindow=false,     
    pdfstartview={FitH},    
    pdftitle={My title},    
    pdfauthor={Author},     
    pdfsubject={Subject},   
    pdfcreator={Creator},   
    pdfproducer={Producer}, 
    pdfkeywords={keyword1} {key2} {key3}, 
    pdfnewwindow=true,      
    colorlinks=true,       
    linkcolor=blue,          
    citecolor=blue,        
    filecolor=blue,      
    urlcolor=blue,       
    plainpages=false        
}

\newcommand{\bk}{{\bf k}}

\begin{document}
\title{Microscopic pairing fingerprint of the iron-based superconductor ${\rm Ba_{1-x}K_xFe_2As_2}$}
\date{\today}
\author{T. B\"ohm}
\affiliation{Walther Meissner Institut, Bayerische Akademie der Wissenschaften, 85748 Garching, Germany}
\affiliation{Fakult\"at f\"ur Physik E23, Technische Universit\"at M\"unchen, 85748 Garching, Germany}
\affiliation{Stanford Institute for Materials and Energy Sciences, SLAC National Accelerator Laboratory, 2575 Sand Hill Road, Menlo Park, California 94025, USA}
\author{F. Kretzschmar}
\altaffiliation{Present address: Intel Mobile Communications, Am Campeon 10-12, 85579 Neubiberg, Germany}
\affiliation{Walther Meissner Institut, Bayerische Akademie der Wissenschaften, 85748 Garching, Germany}
\affiliation{Fakult\"at f\"ur Physik E23, Technische Universit\"at M\"unchen, 85748 Garching, Germany}
\author{A. Baum}
\affiliation{Walther Meissner Institut, Bayerische Akademie der Wissenschaften, 85748 Garching, Germany}
\affiliation{Fakult\"at f\"ur Physik E23, Technische Universit\"at M\"unchen, 85748 Garching, Germany}
\author{M. Rehm}
\altaffiliation{Present address: Schreinerei Kugler, Augsburger Str. 6, 86633 Neuburg, Germany}
\affiliation{Walther Meissner Institut, Bayerische Akademie der Wissenschaften, 85748 Garching, Germany}
\affiliation{Fakult\"at f\"ur Physik E23, Technische Universit\"at M\"unchen, 85748 Garching, Germany}
\author{D. Jost}
\affiliation{Walther Meissner Institut, Bayerische Akademie der Wissenschaften, 85748 Garching, Germany}
\affiliation{Fakult\"at f\"ur Physik E23, Technische Universit\"at M\"unchen, 85748 Garching, Germany}
\author{R.~Hosseinian~Ahangharnejhad}
\altaffiliation{Present address: School of Solar and Advanced Renewable Energy, University of Toledo, Toledo, Ohio 43606, USA}
\affiliation{Walther Meissner Institut, Bayerische Akademie der Wissenschaften, 85748 Garching, Germany}
\affiliation{Fakult\"at f\"ur Physik E23, Technische Universit\"at M\"unchen, 85748 Garching, Germany}
\author{R.~Thomale}
\affiliation{Theoretical Physics, University of W\"urzburg, 97074 W\"urzburg, Germany}
\author{C. Platt}
\affiliation{Department of Physics, McCullough Building, Stanford University, Stanford, California 94305-4045, USA}
\author{T.\,A. Maier}
\affiliation{Center for Nanophase Materials Sciences and Computer Science and Mathematics Division, Oak Ridge National Laboratory, Oak Ridge, Tennessee 37831-6494, USA}
\author{W. Hanke}
\affiliation{Theoretical Physics, University of W\"urzburg, 97074 W\"urzburg, Germany}
\author{B.~Moritz}
\affiliation{Stanford Institute for Materials and Energy Sciences, SLAC National Accelerator Laboratory, 2575 Sand Hill Road, Menlo Park, California 94025, USA}
\author{T.\,P.~Devereaux}
\affiliation{Stanford Institute for Materials and Energy Sciences, SLAC National Accelerator Laboratory, 2575 Sand Hill Road, Menlo Park, California 94025, USA}
\affiliation{Geballe Laboratory for Advanced Materials, Stanford University, Stanford, CA 94305, USA}
\author{D.\,J.~Scalapino}
\affiliation{Physics Department, University of California, Santa Barbara, California 93106-9530, USA}
\author{S.~Maiti}
\affiliation{Department of Physics, University of Florida, Gainesville, Florida 32611, USA}
\author{P.\,J. Hirschfeld}
\affiliation{Department of Physics, University of Florida, Gainesville, Florida 32611, USA}
\author{P. Adelmann}
\affiliation{Karlsruher Institut f\"ur Technologie, Institut f\"ur Festk\"orperphysik, 76021 Karlsruhe, Germany}
\author{T. Wolf}
\affiliation{Karlsruher Institut f\"ur Technologie, Institut f\"ur Festk\"orperphysik, 76021 Karlsruhe, Germany}
\author{Hai-Hu Wen}
\affiliation{National Laboratory of Solid State Microstructures and Department of Physics, Nanjing University, Nanjing 210093, China}
\author{R. Hackl}
\affiliation{Walther Meissner Institut, Bayerische Akademie der Wissenschaften, 85748 Garching, Germany}

\begin{abstract}
  Resolving the microscopic pairing mechanism and its experimental identification in unconventional superconductors is among the most vexing problems of contemporary condensed matter physics. We show that Raman spectroscopy provides an avenue for this quest by probing the structure of the pairing interaction at play in an unconventional superconductor. As we study the spectra of the prototypical Fe-based superconductor ${\rm Ba_{1-x}K_xFe_2As_2}$ for $0.22\le x \le 0.70$ in all symmetry channels, Raman spectroscopy allows us to distill the leading $s$-wave state. In addition, the spectra collected in the $B_{1g}$ symmetry channel reveal the existence of two collective modes which are indicative of the presence of two competing, yet sub-dominant, pairing tendencies of $d_{x^2-y^2}$ symmetry type. A comprehensive functional Renormalization Group (fRG) and random-phase approximation (RPA) study on this compound confirms the presence of the two sub-leading channels, and consistently matches the experimental doping dependence of the related modes. The synopsis of experimental evidence and theoretical modelling supports a spin-fluctuation mediated superconducting pairing mechanism.
\end{abstract}

\pacs{74.20.Mn
, 74.20.Rp
, 74.70.Xa
, 74.25.nd
}
\maketitle


\section{Microscopic Pairing}
In superconductors such as the cuprates, ferro-pnictides,
ruthenates or heavy-fermion systems, the pairing mechanism is
believed to be unconventional and related to direct electronic
interactions rather than conventional electron-phonon mediated
couplings. Yet, the precise microscopic mechanism, the
``glue'' that binds electrons into Cooper pairs, remains elusive. Measurements of  the superconducting ground state alone are insufficient to unambiguously determine whether a superconductor has a conventional or unconventional pairing mechanism. Raman spectroscopy provides the avenue for gathering the missing information in \textit{both dominant and sub-dominant} pairing channels.
\begin{figure}[htbp]
  \centering
  \includegraphics[scale=1]{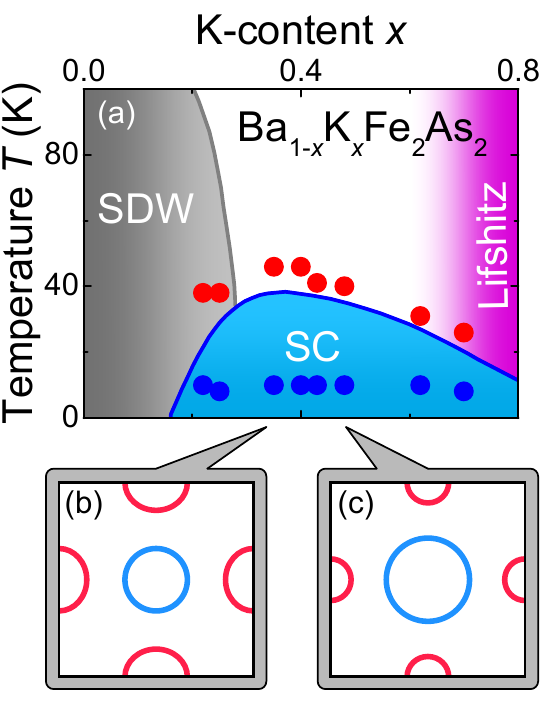}
  \caption{Phase diagram and schematics of doping dependent Fermi surfaces in $\mathrm{Ba_{1-x}K_{x}Fe_2As_2}$. (a) The sampling points of the measurements are compiled in the phase diagram \cite{Bohmer:2014} as blue and red dots deep in the superconducting state and slightly above $T_c$, respectively. (b) and (c) Schematic Brillouin zone and Fermi surface in the 1\,Fe unit cell. With increasing hole-doping $x$ the hole pockets (blue) grow and the electron pockets (red) shrink (changes exaggerated).
  }
  \label{fig:cartoon}
\end{figure}

In comparison to other techniques, Raman spectroscopy (which
involves inelastic scattering of light) is rather unique as it
provides access to both the energy gaps of a superconductor 
and to bound states inside the gaps \cite{Devereaux:2007} that serve
as signposts marking the strength of a given pairing interaction.

These bound states were predicted a long time ago by Bardasis and Schrieffer (BS) \cite{Bardasis:1961} and are collective excitations that correspond to the phase oscillations of the ground state order parameter triggered by the sub-dominant ($d$-wave) interactions. The BS modes or particle-particle excitons couple to the Raman probe, but there is no consensus yet about their observation in conventional superconductors \cite{Monien:1990,Bohm:2014}. Fe-based superconductors (FeSCs), however, presented a more favorable scenario to search for this physics as many of them are believed to exhibit $s_{\pm}$ pairing (with an order parameter that may change sign between Fermi surface pockets \cite{Mazin:2008,Yu:2013,Hosono:2015,Chubukov:2015,Hirschfeld:2016}) and also a sub-leading $d$-wave pairing interaction that can be strongly competitive. Theoretical calculations based on spin fluctuations have even argued that $d$-wave could become the ground state for sufficiently strong hole-doping \cite{Thomale:2009,Thomale:2011a}.

As a result, Scalapino and Devereaux \cite{Scalapino:2009}
performed a `bare-bones' calculation for a typical FeSC
electronic structure with $s_\pm$ symmetry of the ground state
and anisotropic gaps, showing that the mode frequency should
depend on $1/\lambda_d-1/ \lambda_s$, where
$\lambda_d$ and $\lambda_s$  are the respective coupling strengths of the electrons to
the glue that  binds the Cooper pair in the $d$-wave and the $s$-wave channel. Recent measurements on Ba$_{1-x}$K$_x$Fe$_2$As$_2$ \cite{Kretzschmar:2013,Bohm:2014,WuSF:2017},
NaFe$_{1-x}$Co$_x$As \cite{Thorsmolle:2016}, Ba(Fe$_{1-x}$Co$_x$)$_2$As$_2$ \cite{Muschler:2009,Gallais:2016} found peaks in the $B_{1g}$ spectrum which were consistent with a collective mode, but its direct association with a BS mode was unclear.

In this work, we confirm the presence of two sub-dominant
pairing interactions, as predicted theoretically, by
providing an ubiquitous identification of \emph{multiple} BS modes
in the $B_{1g}$ spectrum of the prototypical ferro-pnictide
Ba$_{1-x}$K$_x$Fe$_2$As$_2$ (BKFA). Each sub-dominant pairing interaction results in a BS mode \cite{Maiti:2016}.
This perspective underlies our identification of the two new peaks
in the Raman spectrum with $B_{\rm 1g}$ BS modes. The analysis of our experimental peak energies also supports this scenario and even allows us to empirically extract the relative coupling
strengths $\lambda_{d(1)} / \lambda_s, \lambda_{d(2)} /
\lambda_s$, of the two distinct $B_{1g}$ ($d_{x^2-y^2}$) pairing channels competing with the $s_\pm$ ground state. We could reproduce the presence of all three pairing channels by performing a functional Renormalization Group (fRG) as well as a Random Phase Approximation (RPA) study. Since the fRG calculation includes the leading fluctuations (magnetic, superconducting, charge density wave etc.) whereas the RPA is distinctly based on magnetically driven (i.e. spin-fluctuation-induced) pairing, the agreement of both approaches with each other and the experiment strongly points to a spin-fluctuation scenario in BKFA. Since a direct observation of spin fluctuations below $T_c$ is not achievable by Raman scattering (the relevant scattering states are gapped out) we study the BS modes which remain as the fingerprints of the microscopic pairing interactions.

\section{Results}
\begin{figure*}[htbp]
  \centering
  \includegraphics[scale=1]{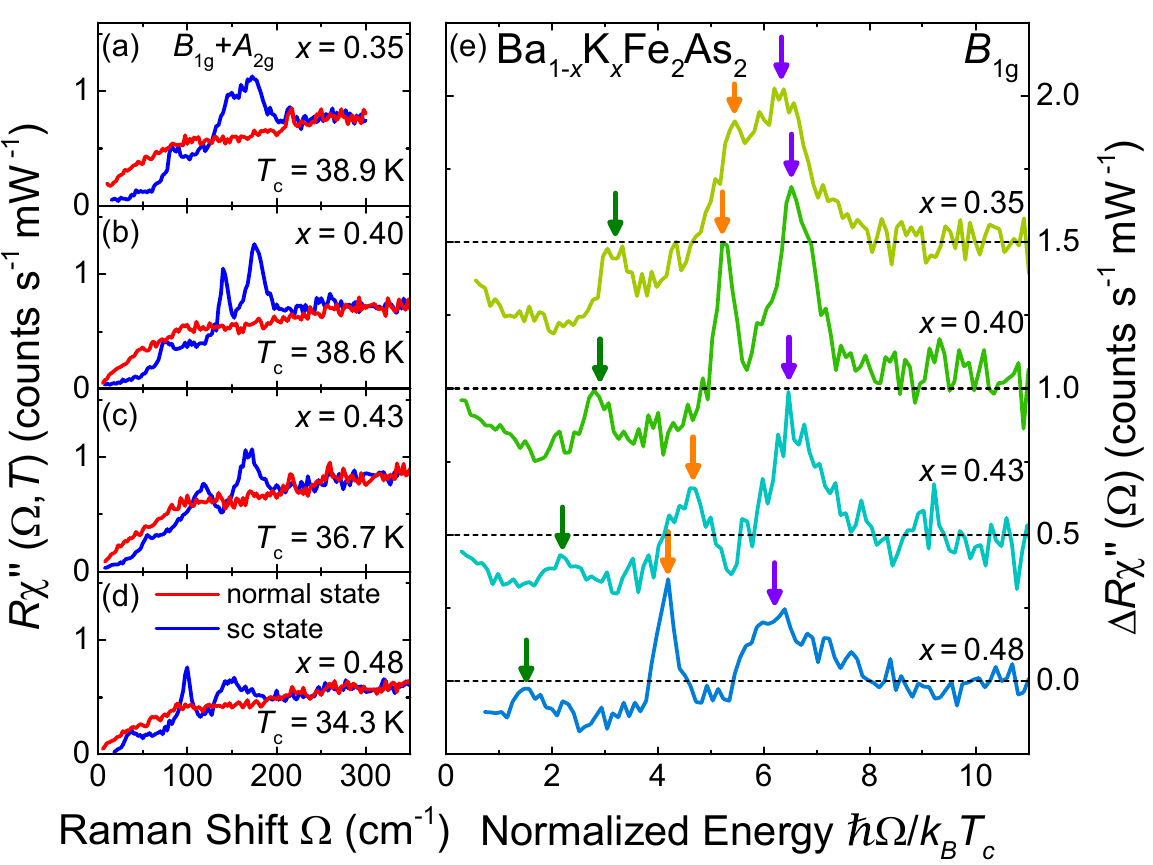}
  \caption{Doping dependence of the Raman spectra in $B_{1g}$ symmetry. (a-d) Raman response $R\chi^{\prime\prime}(\Omega,T,x)$ (raw data after division by the Bose-Einstein factor) of ${\rm Ba_{1-x}K_{x}Fe_2As_2}$ in $B_{1g}$ symmetry above (red) and below (blue) $T_c$ close to optimal doping in the range $x=0.35$ to $x=0.48$. (e) Difference spectra $\Delta R\chi^{\prime\prime}(\Omega,x)$ in $B_{1g}$ symmetry. The energy scale is normalized to the respective $T_c$ values of the differently doped samples. The intensities are off-set, the dashed horizontal lines mark zero. The purple arrows indicate the pair-breaking features at high energy. Green and orange arrows mark two BS modes pulled off the energy gap. They correspond to the sub-dominant channels $d(1)$ and $d(2)$, respectively.}
  \label{fig:B1g-Tc}
\end{figure*}

To this end we measured eight samples of BKFA in the wide doping range $0.22 \leq x \leq 0.70$ as indicated in Fig.~\ref{fig:cartoon}\,(a) and described in detail in Ref.~\onlinecite{Suppl_Bohm:2017}. BKFA forms high quality
single crystals \cite{Rotter:2008,Shen:2011,Karkin:2014} and fairly
clean and isotropic gaps \cite{Evtushinsky:2009,Nakayama:2011}. In the samples with $x=0.22$ and $x=0.25$  superconductivity and the spin density wave (SDW) state coexist. The
samples with $x=0.62$ and $x=0.70$ are close to a
putative Lifshitz transition at $x\sim 0.8$ \cite{Xu:2013}. To
present the case for the physics of sub-dominant pairing interactions,
we wish to stay away from special effects arising from magnetism
or disappearance of pockets and focus on the samples with
$x=0.35,0.40,0.43,0.48$. In this range, the Raman
spectra in the $B_{1g}$ symmetry channel (1\,Fe unit cell) change
continuously as shown in
Fig.~\ref{fig:B1g-Tc}\,(a--d). Spectra of the other symmetries and
outside the range $0.35\le x \le 0.48$ are compiled in Ref.~\onlinecite{Suppl_Bohm:2017}.

The spectra above the superconducting transition temperature $T_c$ are dominated by the electron-hole continua. Below $T_c$ additional (symmetry-dependent) structures appear in the energy range up to approximately 300\,cm$^{-1}$, and the spectral weight is redistributed from below twice the superconducting gap $2\Delta$ to energies above. New features arise from pair breaking, excitations across the gap, and exciton-like bound states \cite{Devereaux:2007,Bohm:2014,Maiti:2016}. With increasing doping and a concomitant reduction of $T_c$, the peaks move to lower energies.

To illustrate why BKFA is a model superconductor for investigating BS modes we highlight the changes in the electronic spectra below $T_c$. For this purpose we subtract the normal state response from the superconducting spectra. This procedure elimantes temperature-independent components of the spectra like the $A_{2g}$ response and phonons in the symmetries $A_{1g}$ and $B_{2g}$ \cite{Suppl_Bohm:2017}. By plotting the difference $\Delta R\chi^{\prime\prime}(\tilde{\Omega}) \equiv R\chi^{\prime\prime}(\tilde{\Omega},T=10\,{\rm K}) - R\chi^{\prime\prime}(\tilde{\Omega},T\gtrsim T_c)$ in Fig.~\ref{fig:B1g-Tc}\,(e) with $\tilde{\Omega} = \hbar\Omega/k_BT_c$ we extract superconductivity-induced features of pure $B_{1g}$ symmetry. Due to the full gap, the difference spectra become negative at low energies and three pronounced peaks are observed. The differences between normal and superconducting spectra disappear ($\Delta R\chi^{\prime\prime}\rightarrow 0$) close to $\tilde{\Omega}=8$. The highest peak [purple arrows in Fig.~\ref{fig:B1g-Tc}\,(e)] at approximately 6.2, which we identify with the maximal gap, depends weakly on doping. The range of $2\Delta/k_BT_c \simeq 6.2$ is in qualitative agreement with the results from other methods \cite{Evtushinsky:2009,Nakayama:2011,Hardy:2016}. This enables us to check the validity of the RPA and fRG approaches in a coupled system of intermediate strength. There are two additional narrow lines in the ranges 1.5--3 (green arrows) and 4--5.5 (orange arrows) displaying a strong monotonic downshift with increasing K content.

At optimal doping ($x=0.40$), evidence was furnished that the narrow line at 5.3 [140\,cm$^{-1}$ in Fig.~\ref{fig:B1g-Tc}\,(b)] results from a bound state of two electrons of a broken Cooper pair \cite{Bohm:2014}. The other narrow line at 2.8 [75\,cm$^{-1}$ in Fig.~\ref{fig:B1g-Tc}\,(b)] is difficult to properly assign on the basis of just one doping level. A suggestion for this peak as a second, smaller pair-breaking peak  and a single BS mode at 5.3 was given in e.g. Ref. \onlinecite{Bohm:2014}. However, upon studying several doping levels and all symmetries \cite{Suppl_Bohm:2017} we find the following systematics in favor of two BS modes: (i) The two in-gap modes appear only in $B_{1g}$ symmetry. (ii) As opposed to the pair-breaking maxima at approximately $6\,k_{\rm B}T_{\rm c}$ there are no other gap energies observed the two sharp modes could correspond to. (iii) Upon doping K for Ba the in-gap modes increasingly split off of the pair-breaking maximum. The nearly identical doping dependences of the two modes and the absence of pair-breaking features in other symmetries suggest that both modes are linked to the maximal gap. The unique appearance of narrow BS modes in $B_{1g}$ symmetry for $0.35\le x \le 0.48$ indicates that there are sub-dominant interactions with $d$-wave symmetry. We label the corresponding sub-leading $B_{1g}$ channels as $d(1)$ and $d(2)$ for the lower- and the higher-energy line, respectively.

\begin{figure}[tbp]
  \centering
  \includegraphics[scale=1]{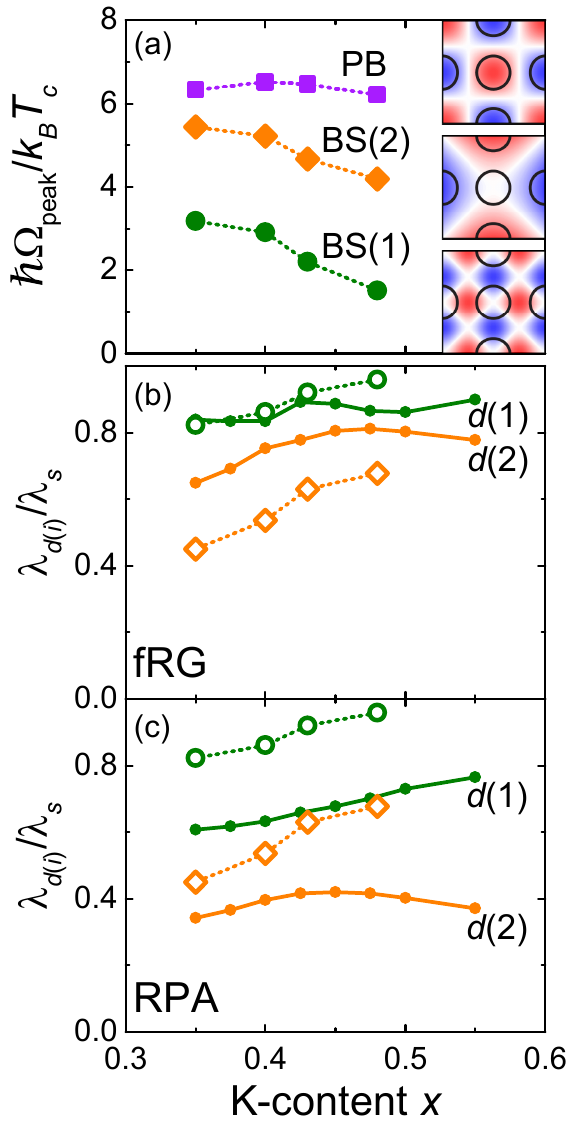}
  \caption{Gap energies and relative coupling strengths. (a) Doping dependence of the characteristic $B_{1g}$ gap energies. The highest pair-breaking energy $\Omega_\mathrm{PB}$ scales approximately with $T_c$. The maxima at $\Omega_\mathrm{BS(1)}$ and $\Omega_\mathrm{BS(2)}$ inside the gap decrease faster than $T_c$. (b) and (c) Relative coupling parameters of the sub-dominant ($\lambda_{d(i)}$) and the dominant ($\lambda_s$) channel. With a dominant interaction of $\lambda_s = 0.7$ \cite{Kuroki:2009a,Ikeda:2010}, the ratios for $\lambda_{d(i)}/\lambda_s$ are extracted from the experiment (open green and orange symbols, corresponding to $d(1)$ and $d(2)$, respectively). The green and orange dots represent results from fRG and RPA \cite{Maiti:2016} calculations in panels (b) and (c), respectively.}
  \label{fig:lambda}
\end{figure}

In Fig.~\ref{fig:lambda}\,(a) we compile experimental peak energies
derived from Fig.~\ref{fig:B1g-Tc}. The difference between $2\Delta$ (purple) and the BS modes in the range $1.5-5\,k_BT_c$
(green and orange) corresponds to the binding energies
$E_{b(i)}=2\Delta-\Omega_{\rm BS(i)}$ with $i=1,\,2$ of the bound
states. The ratios of the relative coupling strengths $\lambda_{d(i)}/\lambda_s$ are estimated from $E_{b(i)}/2\Delta$ using the results of Refs. \onlinecite{Monien:1990,Scalapino:2009,Bohm:2014} and $\lambda_s=0.7$ from Refs. \onlinecite{Ikeda:2010} and \onlinecite{Kuroki:2009a}. Note that we used a doping-independent value of 0.7 for this estimate as the ratios $\lambda_{d(i)}/\lambda_s$ are weakly sensitive to small changes of $\lambda_s$ \cite{Suppl_Bohm:2017}.

\section{Theoretical Methods}
According to Ref.~\onlinecite{Maiti:2016}, the presence of two BS modes in the same symmetry channel must imply the presence of two pairing interactions with different form factors competing  with the ground state. Thus in addition to the ratios $\lambda_{d(i)}/\lambda_s$ derived from experiment, we show in Fig.~\ref{fig:lambda}\,(b) and (c) the results of two microscopic studies using fRG and RPA schemes that precisely identify these pairing channels and also provide an estimate for $\lambda_{d(i)}/\lambda_s$. The comparison of the two independent approaches allows us to pin down the origin of the leading pairing channel since the fRG includes all interactions \cite{Metzner:2012,Platt:2014} whereas the RPA focuses on the spin sector as spelled out in detail in Ref.~\onlinecite{Suppl_Bohm:2017}. Another difference becomes apparent in the procedure used to determine the effective interaction potential. The fRG analysis is designed to start its unbiased renormalization group flow already at energies above the bandwidth while the effective model scale entering the RPA resummation has to be chosen at comparably lower energies\cite{Suppl_Bohm:2017}. As it turns out, however, in spite of these differing intinitalizations, transcending further down to energies at which superconductivity occurs yields similar findings for both methods.

In order to determine the hierarchy of pairing interactions from the effective pairing vertex $V$ from either fRG or RPA, we decompose this pairing channel into eigenmodes, which is tantamount to solving an eigenvalue problem of the form \cite{Suppl_Bohm:2017}
\begin{equation}
  \label{eq:eigenvalue}
  \int_{\text{FS}} dq V(k,q) g_\alpha(q) = \lambda_\alpha g_\alpha(k),
\end{equation}
where $k$ comprises momentum, band, and spin degrees of freedom, and $\alpha$ is the index consecutively numbering the different eigenvalues. We assume $\alpha$ to be ordered according to the magnitude of eigenvalues $\lambda_\alpha$. $g_\alpha(k)$ is the pairing eigenvector along the Fermi surfaces specifying the symmetry of the pairing.

From both fRG and RPA, we find $\lambda_s, g_s(k)$ ($\alpha=1$) to be the dominant superconducting pairing of $A_{1g}$ ($s_\pm$) type and $\lambda_{d(1,2)}, g_{d(1,2)}(k)$ ($\alpha=2,3$) the sub-leading $B_{1g}$ type couplings. Schematic eigenvectors $g_\alpha(\bold{k})$ for $\alpha=1,2,3$ are shown as insets in Fig.~\ref{fig:lambda}\,(a). These results apply to both $V\equiv V^\Lambda_{\text{fRG}}$ and $V\equiv V_{\text{RPA}}$ when used in Eq.~\eqref{eq:eigenvalue}, where $\Lambda$ is the low-energy cutoff in the fRG flow that serves as an upper bound for the transition temperature \cite{Metzner:2012,Platt:2014,Suppl_Bohm:2017}. The leading eigenvalue $\lambda_s \equiv \lambda_1$ in Eq.~\eqref{eq:eigenvalue}, which is a function of $\Lambda$ in the case of fRG, then determines the leading Fermi surface instability. The ratios of the eigenvalues $\lambda_{d(1,2)}/\lambda_s \equiv \lambda_{2,3}/\lambda_1$ determine the peak positions of the BS modes and are shown along with the experiments in Fig.~\ref{fig:lambda}\,(b) and (c). Note that $\lambda_2\equiv \lambda_{d1}$ fits the extended $d$-wave harmonic form predicted in Ref.~\onlinecite{Thomale:2009}.

\section{Conclusions}
From the plethora of theories intended to describe the iron-based superconductors, the comparison with the experiment now enables us, as a first step, to verify the validity of fRG and RPA for the intermediately coupled electronic system of BKFA. We find in accordance with the experiment that both approaches predict the strongest sub-leading channels to be of $d$-wave symmetry. Furthermore, the theoretical predictions for the relative coupling parameters as shown in Fig.~\ref{fig:lambda} are in good agreement with the experiment. The fRG results are in quantitative agreement, the RPA values systematically underestimate the relative coupling strength but are still close to the experiment. Hence we conclude that fRG and RPA are suitable to describe the experiment around optimal doping, $0.35 \le x \le 0.48$, where the two collective BS modes can be identified. Besides the agreement with the experiment the fRG interaction eigenvectors $g_\alpha(\bold{k})$ match very well with those obtained from the spin-fluctuation-based RPA analysis in all three channels ($\alpha=1,2,3$). These agreements indicate that spin fluctuations are an important if not the leading interaction in the system under consideration.

The results presented here put narrow constraints on the description of the Raman data and render differing interpretations \cite{Gallais:2016,Thorsmolle:2016} rather unlikely to be applicable to BKFA. Hence, the observation of two collective modes inside the gap of a superconductor establishes a novelty in terms of experimental analysis which promises to have an impact on the general understanding of unconventional superconductivity. Along with the magnitude of the gap, the modes reveal the hierarchy of pairing states in a prototypical material, in full agreement with microscopic predictions. As a result, our experiment demonstrates the unique possibilities of using light scattering as a probe for observing unconventional pairing fingerprints.

\begin{acknowledgments}
We acknowledge useful discussions with L. Benfatto, A. Eberlein, D. Einzel, S. A. Kivelson, C. Meingast, and I. T\"utt\H{o}. W.H. gratefully acknowledges the hospitality of the Institute for Theoretical Physics at the University of California Santa Barbara. Financial support for the work came from the Deutsche Forschungsgemeinschaft (DFG) via the Priority Program SPP\,1458 (T.B., A.B., R.H., C.P. and W.H., project nos. HA\,2071/7-2 and HA\,1537/24-2), the Collaborative Research Centers SFB\,1170 (W.H., C.P., and R.T.) and TRR\,80 (F.K. and R.H.), the Bavarian Californian Technology Center BaCaTeC (T.B. and R.H., project no. A5\,[2012-2]), the European Research Council (ERC) through ERC-StG-Thomale-TOPOLECTRICS (R.T.), and from the U.S. Department of Energy (DOE), Office of Basic Energy Sciences (B.M. and T.P.D.), Division of Materials Sciences and Engineering, under Contract No. DEAC02-76SF00515. The RPA calculations were conducted at the Center for Nanophase Materials Sciences, which is a DOE Office of Science User Facility. The work in China (H.-H.W.) was supported by the National Key Research and Development Program of China (2016YFA0300401), and the National Natural Science Foundation of China (NSFC) via projects A0402/11534005 and A0402/11374144.
\end{acknowledgments}

\bibliography{Boehm_BKFA_doping_171021}

\begin{thebibliography}{42}%
\makeatletter
\providecommand \@ifxundefined [1]{%
 \@ifx{#1\undefined}
}%
\providecommand \@ifnum [1]{%
 \ifnum #1\expandafter \@firstoftwo
 \else \expandafter \@secondoftwo
 \fi
}%
\providecommand \@ifx [1]{%
 \ifx #1\expandafter \@firstoftwo
 \else \expandafter \@secondoftwo
 \fi
}%
\providecommand \natexlab [1]{#1}%
\providecommand \enquote  [1]{``#1''}%
\providecommand \bibnamefont  [1]{#1}%
\providecommand \bibfnamefont [1]{#1}%
\providecommand \citenamefont [1]{#1}%
\providecommand \href@noop [0]{\@secondoftwo}%
\providecommand \href [0]{\begingroup \@sanitize@url \@href}%
\providecommand \@href[1]{\@@startlink{#1}\@@href}%
\providecommand \@@href[1]{\endgroup#1\@@endlink}%
\providecommand \@sanitize@url [0]{\catcode `\\12\catcode `\$12\catcode
  `\&12\catcode `\#12\catcode `\^12\catcode `\_12\catcode `\%12\relax}%
\providecommand \@@startlink[1]{}%
\providecommand \@@endlink[0]{}%
\providecommand \url  [0]{\begingroup\@sanitize@url \@url }%
\providecommand \@url [1]{\endgroup\@href {#1}{\urlprefix }}%
\providecommand \urlprefix  [0]{URL }%
\providecommand \Eprint [0]{\href }%
\providecommand \doibase [0]{http://dx.doi.org/}%
\providecommand \selectlanguage [0]{\@gobble}%
\providecommand \bibinfo  [0]{\@secondoftwo}%
\providecommand \bibfield  [0]{\@secondoftwo}%
\providecommand \translation [1]{[#1]}%
\providecommand \BibitemOpen [0]{}%
\providecommand \bibitemStop [0]{}%
\providecommand \bibitemNoStop [0]{.\EOS\space}%
\providecommand \EOS [0]{\spacefactor3000\relax}%
\providecommand \BibitemShut  [1]{\csname bibitem#1\endcsname}%
\let\auto@bib@innerbib\@empty
\bibitem [{\citenamefont {B{\"o}hmer}\ \emph {et~al.}(2014)\citenamefont
  {B{\"o}hmer}, \citenamefont {Burger}, \citenamefont {Hardy}, \citenamefont
  {Wolf}, \citenamefont {Schweiss}, \citenamefont {Fromknecht}, \citenamefont
  {Reinecker}, \citenamefont {Schranz},\ and\ \citenamefont
  {Meingast}}]{Bohmer:2014}%
  \BibitemOpen
  \bibfield  {author} {\bibinfo {author} {\bibfnamefont {A.~E.}\ \bibnamefont
  {B{\"o}hmer}}, \bibinfo {author} {\bibfnamefont {P.}~\bibnamefont {Burger}},
  \bibinfo {author} {\bibfnamefont {F.}~\bibnamefont {Hardy}}, \bibinfo
  {author} {\bibfnamefont {T.}~\bibnamefont {Wolf}}, \bibinfo {author}
  {\bibfnamefont {P.}~\bibnamefont {Schweiss}}, \bibinfo {author}
  {\bibfnamefont {R.}~\bibnamefont {Fromknecht}}, \bibinfo {author}
  {\bibfnamefont {M.}~\bibnamefont {Reinecker}}, \bibinfo {author}
  {\bibfnamefont {W.}~\bibnamefont {Schranz}}, \ and\ \bibinfo {author}
  {\bibfnamefont {C.}~\bibnamefont {Meingast}},\ }\href {\doibase
  10.1103/PhysRevLett.112.047001} {\bibfield  {journal} {\bibinfo  {journal}
  {Phys. Rev. Lett.}\ }\textbf {\bibinfo {volume} {112}},\ \bibinfo {pages}
  {047001} (\bibinfo {year} {2014})}\BibitemShut {NoStop}%
\bibitem [{\citenamefont {Devereaux}\ and\ \citenamefont
  {Hackl}(2007)}]{Devereaux:2007}%
  \BibitemOpen
  \bibfield  {author} {\bibinfo {author} {\bibfnamefont {T.~P.}\ \bibnamefont
  {Devereaux}}\ and\ \bibinfo {author} {\bibfnamefont {R.}~\bibnamefont
  {Hackl}},\ }\href {\doibase 10.1103/RevModPhys.79.175} {\bibfield  {journal}
  {\bibinfo  {journal} {Rev. Mod. Phys.}\ }\textbf {\bibinfo {volume} {79}},\
  \bibinfo {pages} {175} (\bibinfo {year} {2007})}\BibitemShut {NoStop}%
\bibitem [{\citenamefont {Bardasis}\ and\ \citenamefont
  {Schrieffer}(1961)}]{Bardasis:1961}%
  \BibitemOpen
  \bibfield  {author} {\bibinfo {author} {\bibfnamefont {A.}~\bibnamefont
  {Bardasis}}\ and\ \bibinfo {author} {\bibfnamefont {J.~R.}\ \bibnamefont
  {Schrieffer}},\ }\href {\doibase 10.1103/PhysRev.121.1050} {\bibfield
  {journal} {\bibinfo  {journal} {Phys. Rev.}\ }\textbf {\bibinfo {volume}
  {121}},\ \bibinfo {pages} {1050} (\bibinfo {year} {1961})}\BibitemShut
  {NoStop}%
\bibitem [{\citenamefont {Monien}\ and\ \citenamefont
  {Zawadowski}(1990)}]{Monien:1990}%
  \BibitemOpen
  \bibfield  {author} {\bibinfo {author} {\bibfnamefont {H.}~\bibnamefont
  {Monien}}\ and\ \bibinfo {author} {\bibfnamefont {A.}~\bibnamefont
  {Zawadowski}},\ }\href {\doibase 10.1103/PhysRevB.41.8798} {\bibfield
  {journal} {\bibinfo  {journal} {Phys. Rev. B}\ }\textbf {\bibinfo {volume}
  {41}},\ \bibinfo {pages} {8798} (\bibinfo {year} {1990})}\BibitemShut
  {NoStop}%
\bibitem [{\citenamefont {B\"ohm}\ \emph {et~al.}(2014)\citenamefont {B\"ohm},
  \citenamefont {Kemper}, \citenamefont {Moritz}, \citenamefont {Kretzschmar},
  \citenamefont {Muschler}, \citenamefont {Eiter}, \citenamefont {Hackl},
  \citenamefont {Devereaux}, \citenamefont {Scalapino},\ and\ \citenamefont
  {Wen}}]{Bohm:2014}%
  \BibitemOpen
  \bibfield  {author} {\bibinfo {author} {\bibfnamefont {T.}~\bibnamefont
  {B\"ohm}}, \bibinfo {author} {\bibfnamefont {A.~F.}\ \bibnamefont {Kemper}},
  \bibinfo {author} {\bibfnamefont {B.}~\bibnamefont {Moritz}}, \bibinfo
  {author} {\bibfnamefont {F.}~\bibnamefont {Kretzschmar}}, \bibinfo {author}
  {\bibfnamefont {B.}~\bibnamefont {Muschler}}, \bibinfo {author}
  {\bibfnamefont {H.-M.}\ \bibnamefont {Eiter}}, \bibinfo {author}
  {\bibfnamefont {R.}~\bibnamefont {Hackl}}, \bibinfo {author} {\bibfnamefont
  {T.~P.}\ \bibnamefont {Devereaux}}, \bibinfo {author} {\bibfnamefont {D.~J.}\
  \bibnamefont {Scalapino}}, \ and\ \bibinfo {author} {\bibfnamefont {H.-H.}\
  \bibnamefont {Wen}},\ }\href {\doibase 10.1103/PhysRevX.4.041046} {\bibfield
  {journal} {\bibinfo  {journal} {Phys. Rev. X}\ }\textbf {\bibinfo {volume}
  {4}},\ \bibinfo {pages} {041046} (\bibinfo {year} {2014})}\BibitemShut
  {NoStop}%
\bibitem [{\citenamefont {Mazin}\ \emph {et~al.}(2008)\citenamefont {Mazin},
  \citenamefont {Singh}, \citenamefont {Johannes},\ and\ \citenamefont
  {Du}}]{Mazin:2008}%
  \BibitemOpen
  \bibfield  {author} {\bibinfo {author} {\bibfnamefont {I.~I.}\ \bibnamefont
  {Mazin}}, \bibinfo {author} {\bibfnamefont {D.~J.}\ \bibnamefont {Singh}},
  \bibinfo {author} {\bibfnamefont {M.~D.}\ \bibnamefont {Johannes}}, \ and\
  \bibinfo {author} {\bibfnamefont {M.~H.}\ \bibnamefont {Du}},\ }\href
  {\doibase 10.1103/PhysRevLett.101.057003} {\bibfield  {journal} {\bibinfo
  {journal} {Phys. Rev. Lett.}\ }\textbf {\bibinfo {volume} {101}},\ \bibinfo
  {eid} {057003} (\bibinfo {year} {2008})}\BibitemShut {NoStop}%
\bibitem [{\citenamefont {Yu}\ \emph {et~al.}(2013)\citenamefont {Yu},
  \citenamefont {Si}, \citenamefont {Goswami},\ and\ \citenamefont
  {Abrahams}}]{Yu:2013}%
  \BibitemOpen
  \bibfield  {author} {\bibinfo {author} {\bibfnamefont {R.}~\bibnamefont
  {Yu}}, \bibinfo {author} {\bibfnamefont {Q.}~\bibnamefont {Si}}, \bibinfo
  {author} {\bibfnamefont {P.}~\bibnamefont {Goswami}}, \ and\ \bibinfo
  {author} {\bibfnamefont {E.}~\bibnamefont {Abrahams}},\ }\href
  {http://stacks.iop.org/1742-6596/449/i=1/a=012025} {\bibfield  {journal}
  {\bibinfo  {journal} {Journal of Physics: Conference Series}\ }\textbf
  {\bibinfo {volume} {449}},\ \bibinfo {pages} {012025} (\bibinfo {year}
  {2013})}\BibitemShut {NoStop}%
\bibitem [{\citenamefont {Hosono}\ and\ \citenamefont
  {Kuroki}(2015)}]{Hosono:2015}%
  \BibitemOpen
  \bibfield  {author} {\bibinfo {author} {\bibfnamefont {H.}~\bibnamefont
  {Hosono}}\ and\ \bibinfo {author} {\bibfnamefont {K.}~\bibnamefont
  {Kuroki}},\ }\href {\doibase http://dx.doi.org/10.1016/j.physc.2015.02.020}
  {\bibfield  {journal} {\bibinfo  {journal} {Physica C: Superconductivity and
  its Applications}\ }\textbf {\bibinfo {volume} {514}},\ \bibinfo {pages} {399
  } (\bibinfo {year} {2015})},\ \bibinfo {note} {superconducting Materials:
  Conventional, Unconventional and Undetermined}\BibitemShut {NoStop}%
\bibitem [{\citenamefont {Chubukov}\ \emph {et~al.}(2015)\citenamefont
  {Chubukov}, \citenamefont {Fernandes},\ and\ \citenamefont
  {Schmalian}}]{Chubukov:2015}%
  \BibitemOpen
  \bibfield  {author} {\bibinfo {author} {\bibfnamefont {A.~V.}\ \bibnamefont
  {Chubukov}}, \bibinfo {author} {\bibfnamefont {R.~M.}\ \bibnamefont
  {Fernandes}}, \ and\ \bibinfo {author} {\bibfnamefont {J.}~\bibnamefont
  {Schmalian}},\ }\href {\doibase 10.1103/PhysRevB.91.201105} {\bibfield
  {journal} {\bibinfo  {journal} {Phys. Rev. B}\ }\textbf {\bibinfo {volume}
  {91}},\ \bibinfo {pages} {201105} (\bibinfo {year} {2015})}\BibitemShut
  {NoStop}%
\bibitem [{\citenamefont {Hirschfeld}(2016)}]{Hirschfeld:2016}%
  \BibitemOpen
  \bibfield  {author} {\bibinfo {author} {\bibfnamefont {P.~J.}\ \bibnamefont
  {Hirschfeld}},\ }\href {\doibase
  http://dx.doi.org/10.1016/j.crhy.2015.10.002} {\bibfield  {journal} {\bibinfo
   {journal} {C. R. Physique}\ }\textbf {\bibinfo {volume} {17}},\ \bibinfo
  {pages} {197 } (\bibinfo {year} {2016})},\ \bibinfo {note} {iron-based
  superconductors / Supraconducteurs � base de fer}\BibitemShut {NoStop}%
\bibitem [{\citenamefont {Thomale}\ \emph {et~al.}(2009)\citenamefont
  {Thomale}, \citenamefont {Platt}, \citenamefont {Hu}, \citenamefont
  {Honerkamp},\ and\ \citenamefont {Bernevig}}]{Thomale:2009}%
  \BibitemOpen
  \bibfield  {author} {\bibinfo {author} {\bibfnamefont {R.}~\bibnamefont
  {Thomale}}, \bibinfo {author} {\bibfnamefont {C.}~\bibnamefont {Platt}},
  \bibinfo {author} {\bibfnamefont {J.}~\bibnamefont {Hu}}, \bibinfo {author}
  {\bibfnamefont {C.}~\bibnamefont {Honerkamp}}, \ and\ \bibinfo {author}
  {\bibfnamefont {B.~A.}\ \bibnamefont {Bernevig}},\ }\href {\doibase
  10.1103/PhysRevB.80.180505} {\bibfield  {journal} {\bibinfo  {journal} {Phys.
  Rev. B}\ }\textbf {\bibinfo {volume} {80}},\ \bibinfo {pages} {180505}
  (\bibinfo {year} {2009})}\BibitemShut {NoStop}%
\bibitem [{\citenamefont {Thomale}\ \emph {et~al.}(2011)\citenamefont
  {Thomale}, \citenamefont {Platt}, \citenamefont {Hanke}, \citenamefont {Hu},\
  and\ \citenamefont {Bernevig}}]{Thomale:2011a}%
  \BibitemOpen
  \bibfield  {author} {\bibinfo {author} {\bibfnamefont {R.}~\bibnamefont
  {Thomale}}, \bibinfo {author} {\bibfnamefont {C.}~\bibnamefont {Platt}},
  \bibinfo {author} {\bibfnamefont {W.}~\bibnamefont {Hanke}}, \bibinfo
  {author} {\bibfnamefont {J.}~\bibnamefont {Hu}}, \ and\ \bibinfo {author}
  {\bibfnamefont {B.~A.}\ \bibnamefont {Bernevig}},\ }\href {\doibase
  10.1103/PhysRevLett.107.117001} {\bibfield  {journal} {\bibinfo  {journal}
  {Phys. Rev. Lett.}\ }\textbf {\bibinfo {volume} {107}},\ \bibinfo {pages}
  {117001} (\bibinfo {year} {2011})}\BibitemShut {NoStop}%
\bibitem [{\citenamefont {Scalapino}\ and\ \citenamefont
  {Devereaux}(2009)}]{Scalapino:2009}%
  \BibitemOpen
  \bibfield  {author} {\bibinfo {author} {\bibfnamefont {D.~J.}\ \bibnamefont
  {Scalapino}}\ and\ \bibinfo {author} {\bibfnamefont {T.~P.}\ \bibnamefont
  {Devereaux}},\ }\href {\doibase 10.1103/PhysRevB.80.140512} {\bibfield
  {journal} {\bibinfo  {journal} {Phys. Rev. B}\ }\textbf {\bibinfo {volume}
  {80}},\ \bibinfo {eid} {140512} (\bibinfo {year} {2009})}\BibitemShut
  {NoStop}%
\bibitem [{\citenamefont {Kretzschmar}\ \emph {et~al.}(2013)\citenamefont
  {Kretzschmar}, \citenamefont {Muschler}, \citenamefont {B\"ohm},
  \citenamefont {Baum}, \citenamefont {Hackl}, \citenamefont {Wen},
  \citenamefont {Tsurkan}, \citenamefont {Deisenhofer},\ and\ \citenamefont
  {Loidl}}]{Kretzschmar:2013}%
  \BibitemOpen
  \bibfield  {author} {\bibinfo {author} {\bibfnamefont {F.}~\bibnamefont
  {Kretzschmar}}, \bibinfo {author} {\bibfnamefont {B.}~\bibnamefont
  {Muschler}}, \bibinfo {author} {\bibfnamefont {T.}~\bibnamefont {B\"ohm}},
  \bibinfo {author} {\bibfnamefont {A.}~\bibnamefont {Baum}}, \bibinfo {author}
  {\bibfnamefont {R.}~\bibnamefont {Hackl}}, \bibinfo {author} {\bibfnamefont
  {H.-H.}\ \bibnamefont {Wen}}, \bibinfo {author} {\bibfnamefont
  {V.}~\bibnamefont {Tsurkan}}, \bibinfo {author} {\bibfnamefont
  {J.}~\bibnamefont {Deisenhofer}}, \ and\ \bibinfo {author} {\bibfnamefont
  {A.}~\bibnamefont {Loidl}},\ }\href {\doibase 10.1103/PhysRevLett.110.187002}
  {\bibfield  {journal} {\bibinfo  {journal} {Phys. Rev. Lett.}\ }\textbf
  {\bibinfo {volume} {110}},\ \bibinfo {pages} {187002} (\bibinfo {year}
  {2013})}\BibitemShut {NoStop}%
\bibitem [{\citenamefont {Wu}\ \emph {et~al.}(2017)\citenamefont {Wu},
  \citenamefont {Richard}, \citenamefont {Ding}, \citenamefont {Wen},
  \citenamefont {Tan}, \citenamefont {Wang}, \citenamefont {Zhang},
  \citenamefont {Dai},\ and\ \citenamefont {Blumberg}}]{WuSF:2017}%
  \BibitemOpen
  \bibfield  {author} {\bibinfo {author} {\bibfnamefont {S.-F.}\ \bibnamefont
  {Wu}}, \bibinfo {author} {\bibfnamefont {P.}~\bibnamefont {Richard}},
  \bibinfo {author} {\bibfnamefont {H.}~\bibnamefont {Ding}}, \bibinfo {author}
  {\bibfnamefont {H.-H.}\ \bibnamefont {Wen}}, \bibinfo {author} {\bibfnamefont
  {G.}~\bibnamefont {Tan}}, \bibinfo {author} {\bibfnamefont {M.}~\bibnamefont
  {Wang}}, \bibinfo {author} {\bibfnamefont {C.}~\bibnamefont {Zhang}},
  \bibinfo {author} {\bibfnamefont {P.}~\bibnamefont {Dai}}, \ and\ \bibinfo
  {author} {\bibfnamefont {G.}~\bibnamefont {Blumberg}},\ }\href {\doibase
  10.1103/PhysRevB.95.085125} {\bibfield  {journal} {\bibinfo  {journal} {Phys.
  Rev. B}\ }\textbf {\bibinfo {volume} {95}},\ \bibinfo {pages} {085125}
  (\bibinfo {year} {2017})}\BibitemShut {NoStop}%
\bibitem [{\citenamefont {Thorsm\o{}lle}\ \emph {et~al.}(2016)\citenamefont
  {Thorsm\o{}lle}, \citenamefont {Khodas}, \citenamefont {Yin}, \citenamefont
  {Zhang}, \citenamefont {Carr}, \citenamefont {Dai},\ and\ \citenamefont
  {Blumberg}}]{Thorsmolle:2016}%
  \BibitemOpen
  \bibfield  {author} {\bibinfo {author} {\bibfnamefont {V.~K.}\ \bibnamefont
  {Thorsm\o{}lle}}, \bibinfo {author} {\bibfnamefont {M.}~\bibnamefont
  {Khodas}}, \bibinfo {author} {\bibfnamefont {Z.~P.}\ \bibnamefont {Yin}},
  \bibinfo {author} {\bibfnamefont {C.}~\bibnamefont {Zhang}}, \bibinfo
  {author} {\bibfnamefont {S.~V.}\ \bibnamefont {Carr}}, \bibinfo {author}
  {\bibfnamefont {P.}~\bibnamefont {Dai}}, \ and\ \bibinfo {author}
  {\bibfnamefont {G.}~\bibnamefont {Blumberg}},\ }\href {\doibase
  10.1103/PhysRevB.93.054515} {\bibfield  {journal} {\bibinfo  {journal} {Phys.
  Rev. B}\ }\textbf {\bibinfo {volume} {93}},\ \bibinfo {pages} {054515}
  (\bibinfo {year} {2016})}\BibitemShut {NoStop}%
\bibitem [{\citenamefont {Muschler}\ \emph {et~al.}(2009)\citenamefont
  {Muschler}, \citenamefont {Prestel}, \citenamefont {Hackl}, \citenamefont
  {Devereaux}, \citenamefont {Analytis}, \citenamefont {Chu},\ and\
  \citenamefont {Fisher}}]{Muschler:2009}%
  \BibitemOpen
  \bibfield  {author} {\bibinfo {author} {\bibfnamefont {B.}~\bibnamefont
  {Muschler}}, \bibinfo {author} {\bibfnamefont {W.}~\bibnamefont {Prestel}},
  \bibinfo {author} {\bibfnamefont {R.}~\bibnamefont {Hackl}}, \bibinfo
  {author} {\bibfnamefont {T.~P.}\ \bibnamefont {Devereaux}}, \bibinfo {author}
  {\bibfnamefont {J.~G.}\ \bibnamefont {Analytis}}, \bibinfo {author}
  {\bibfnamefont {J.-H.}\ \bibnamefont {Chu}}, \ and\ \bibinfo {author}
  {\bibfnamefont {I.~R.}\ \bibnamefont {Fisher}},\ }\href {\doibase
  10.1103/PhysRevB.80.180510} {\bibfield  {journal} {\bibinfo  {journal} {Phys.
  Rev. B}\ }\textbf {\bibinfo {volume} {80}},\ \bibinfo {pages} {180510}
  (\bibinfo {year} {2009})}\BibitemShut {NoStop}%
\bibitem [{\citenamefont {Gallais}\ \emph {et~al.}(2016)\citenamefont
  {Gallais}, \citenamefont {Paul}, \citenamefont {Chauvi\`ere},\ and\
  \citenamefont {Schmalian}}]{Gallais:2016}%
  \BibitemOpen
  \bibfield  {author} {\bibinfo {author} {\bibfnamefont {Y.}~\bibnamefont
  {Gallais}}, \bibinfo {author} {\bibfnamefont {I.}~\bibnamefont {Paul}},
  \bibinfo {author} {\bibfnamefont {L.}~\bibnamefont {Chauvi\`ere}}, \ and\
  \bibinfo {author} {\bibfnamefont {J.}~\bibnamefont {Schmalian}},\ }\href
  {\doibase 10.1103/PhysRevLett.116.017001} {\bibfield  {journal} {\bibinfo
  {journal} {Phys. Rev. Lett.}\ }\textbf {\bibinfo {volume} {116}},\ \bibinfo
  {pages} {017001} (\bibinfo {year} {2016})}\BibitemShut {NoStop}%
\bibitem [{\citenamefont {Maiti}\ \emph {et~al.}(2016)\citenamefont {Maiti},
  \citenamefont {Maier}, \citenamefont {B\"ohm}, \citenamefont {Hackl},\ and\
  \citenamefont {Hirschfeld}}]{Maiti:2016}%
  \BibitemOpen
  \bibfield  {author} {\bibinfo {author} {\bibfnamefont {S.}~\bibnamefont
  {Maiti}}, \bibinfo {author} {\bibfnamefont {T.~A.}\ \bibnamefont {Maier}},
  \bibinfo {author} {\bibfnamefont {T.}~\bibnamefont {B\"ohm}}, \bibinfo
  {author} {\bibfnamefont {R.}~\bibnamefont {Hackl}}, \ and\ \bibinfo {author}
  {\bibfnamefont {P.~J.}\ \bibnamefont {Hirschfeld}},\ }\href {\doibase
  10.1103/PhysRevLett.117.257001} {\bibfield  {journal} {\bibinfo  {journal}
  {Phys. Rev. Lett.}\ }\textbf {\bibinfo {volume} {117}},\ \bibinfo {pages}
  {257001} (\bibinfo {year} {2016})}\BibitemShut {NoStop}%
\bibitem [{Sup()}]{Suppl_Bohm:2017}%
  \BibitemOpen
  \href@noop {} {}\bibinfo {note} {For details see Appendix.}\BibitemShut
  {Stop}%
\bibitem [{\citenamefont {Rotter}\ \emph {et~al.}(2008)\citenamefont {Rotter},
  \citenamefont {Tegel},\ and\ \citenamefont {Johrendt}}]{Rotter:2008}%
  \BibitemOpen
  \bibfield  {author} {\bibinfo {author} {\bibfnamefont {M.}~\bibnamefont
  {Rotter}}, \bibinfo {author} {\bibfnamefont {M.}~\bibnamefont {Tegel}}, \
  and\ \bibinfo {author} {\bibfnamefont {D.}~\bibnamefont {Johrendt}},\ }\href
  {\doibase 10.1103/PhysRevLett.101.107006} {\bibfield  {journal} {\bibinfo
  {journal} {Phys. Rev. Lett.}\ }\textbf {\bibinfo {volume} {101}},\ \bibinfo
  {eid} {107006} (\bibinfo {year} {2008})}\BibitemShut {NoStop}%
\bibitem [{\citenamefont {Shen}\ \emph {et~al.}(2011)\citenamefont {Shen},
  \citenamefont {Yang}, \citenamefont {Wang}, \citenamefont {Han},
  \citenamefont {Zeng}, \citenamefont {Shan}, \citenamefont {Cong},\ and\
  \citenamefont {Wen}}]{Shen:2011}%
  \BibitemOpen
  \bibfield  {author} {\bibinfo {author} {\bibfnamefont {B.}~\bibnamefont
  {Shen}}, \bibinfo {author} {\bibfnamefont {H.}~\bibnamefont {Yang}}, \bibinfo
  {author} {\bibfnamefont {Z.-S.}\ \bibnamefont {Wang}}, \bibinfo {author}
  {\bibfnamefont {F.}~\bibnamefont {Han}}, \bibinfo {author} {\bibfnamefont
  {B.}~\bibnamefont {Zeng}}, \bibinfo {author} {\bibfnamefont {L.}~\bibnamefont
  {Shan}}, \bibinfo {author} {\bibfnamefont {R.}~\bibnamefont {Cong}}, \ and\
  \bibinfo {author} {\bibfnamefont {H.-H.}\ \bibnamefont {Wen}},\ }\href
  {\doibase 10.1103/PhysRevB.84.184512} {\bibfield  {journal} {\bibinfo
  {journal} {Phys. Rev. B}\ }\textbf {\bibinfo {volume} {84}},\ \bibinfo
  {pages} {184512} (\bibinfo {year} {2011})}\BibitemShut {NoStop}%
\bibitem [{\citenamefont {Karkin}\ \emph {et~al.}(2014)\citenamefont {Karkin},
  \citenamefont {Wolf},\ and\ \citenamefont {Goshchitskii}}]{Karkin:2014}%
  \BibitemOpen
  \bibfield  {author} {\bibinfo {author} {\bibfnamefont {A.~E.}\ \bibnamefont
  {Karkin}}, \bibinfo {author} {\bibfnamefont {T.}~\bibnamefont {Wolf}}, \ and\
  \bibinfo {author} {\bibfnamefont {B.~N.}\ \bibnamefont {Goshchitskii}},\
  }\href {\doibase 10.1088/0953-8984/26/27/275702} {\bibfield  {journal}
  {\bibinfo  {journal} {J. Phys.: Condens. Matter}\ }\textbf {\bibinfo {volume}
  {26}},\ \bibinfo {pages} {275702} (\bibinfo {year} {2014})}\BibitemShut
  {NoStop}%
\bibitem [{\citenamefont {Evtushinsky}\ \emph {et~al.}(2009)\citenamefont
  {Evtushinsky}, \citenamefont {Inosov}, \citenamefont {Zabolotnyy},
  \citenamefont {Koitzsch}, \citenamefont {Knupfer}, \citenamefont {B\"uchner},
  \citenamefont {Viazovska}, \citenamefont {Sun}, \citenamefont {Hinkov},
  \citenamefont {Boris}, \citenamefont {Lin}, \citenamefont {Keimer},
  \citenamefont {Varykhalov}, \citenamefont {Kordyuk},\ and\ \citenamefont
  {Borisenko}}]{Evtushinsky:2009}%
  \BibitemOpen
  \bibfield  {author} {\bibinfo {author} {\bibfnamefont {D.~V.}\ \bibnamefont
  {Evtushinsky}}, \bibinfo {author} {\bibfnamefont {D.~S.}\ \bibnamefont
  {Inosov}}, \bibinfo {author} {\bibfnamefont {V.~B.}\ \bibnamefont
  {Zabolotnyy}}, \bibinfo {author} {\bibfnamefont {A.}~\bibnamefont
  {Koitzsch}}, \bibinfo {author} {\bibfnamefont {M.}~\bibnamefont {Knupfer}},
  \bibinfo {author} {\bibfnamefont {B.}~\bibnamefont {B\"uchner}}, \bibinfo
  {author} {\bibfnamefont {M.~S.}\ \bibnamefont {Viazovska}}, \bibinfo {author}
  {\bibfnamefont {G.~L.}\ \bibnamefont {Sun}}, \bibinfo {author} {\bibfnamefont
  {V.}~\bibnamefont {Hinkov}}, \bibinfo {author} {\bibfnamefont {A.~V.}\
  \bibnamefont {Boris}}, \bibinfo {author} {\bibfnamefont {C.~T.}\ \bibnamefont
  {Lin}}, \bibinfo {author} {\bibfnamefont {B.}~\bibnamefont {Keimer}},
  \bibinfo {author} {\bibfnamefont {A.}~\bibnamefont {Varykhalov}}, \bibinfo
  {author} {\bibfnamefont {A.~A.}\ \bibnamefont {Kordyuk}}, \ and\ \bibinfo
  {author} {\bibfnamefont {S.~V.}\ \bibnamefont {Borisenko}},\ }\href {\doibase
  10.1103/PhysRevB.79.054517} {\bibfield  {journal} {\bibinfo  {journal} {Phys.
  Rev. B}\ }\textbf {\bibinfo {volume} {79}},\ \bibinfo {eid} {054517}
  (\bibinfo {year} {2009})}\BibitemShut {NoStop}%
\bibitem [{\citenamefont {Nakayama}\ \emph {et~al.}(2011)\citenamefont
  {Nakayama}, \citenamefont {Sato}, \citenamefont {Richard}, \citenamefont
  {Xu}, \citenamefont {Kawahara}, \citenamefont {Umezawa}, \citenamefont
  {Qian}, \citenamefont {Neupane}, \citenamefont {Chen}, \citenamefont {Ding},\
  and\ \citenamefont {Takahashi}}]{Nakayama:2011}%
  \BibitemOpen
  \bibfield  {author} {\bibinfo {author} {\bibfnamefont {K.}~\bibnamefont
  {Nakayama}}, \bibinfo {author} {\bibfnamefont {T.}~\bibnamefont {Sato}},
  \bibinfo {author} {\bibfnamefont {P.}~\bibnamefont {Richard}}, \bibinfo
  {author} {\bibfnamefont {Y.-M.}\ \bibnamefont {Xu}}, \bibinfo {author}
  {\bibfnamefont {T.}~\bibnamefont {Kawahara}}, \bibinfo {author}
  {\bibfnamefont {K.}~\bibnamefont {Umezawa}}, \bibinfo {author} {\bibfnamefont
  {T.}~\bibnamefont {Qian}}, \bibinfo {author} {\bibfnamefont {M.}~\bibnamefont
  {Neupane}}, \bibinfo {author} {\bibfnamefont {G.~F.}\ \bibnamefont {Chen}},
  \bibinfo {author} {\bibfnamefont {H.}~\bibnamefont {Ding}}, \ and\ \bibinfo
  {author} {\bibfnamefont {T.}~\bibnamefont {Takahashi}},\ }\href {\doibase
  10.1103/PhysRevB.83.020501} {\bibfield  {journal} {\bibinfo  {journal} {Phys.
  Rev. B}\ }\textbf {\bibinfo {volume} {83}},\ \bibinfo {pages} {020501}
  (\bibinfo {year} {2011})}\BibitemShut {NoStop}%
\bibitem [{\citenamefont {Xu}\ \emph {et~al.}(2013)\citenamefont {Xu},
  \citenamefont {Richard}, \citenamefont {Shi}, \citenamefont {van Roekeghem},
  \citenamefont {Qian}, \citenamefont {Razzoli}, \citenamefont {Rienks},
  \citenamefont {Chen}, \citenamefont {Ieki}, \citenamefont {Nakayama},
  \citenamefont {Sato}, \citenamefont {Takahashi}, \citenamefont {Shi},\ and\
  \citenamefont {Ding}}]{Xu:2013}%
  \BibitemOpen
  \bibfield  {author} {\bibinfo {author} {\bibfnamefont {N.}~\bibnamefont
  {Xu}}, \bibinfo {author} {\bibfnamefont {P.}~\bibnamefont {Richard}},
  \bibinfo {author} {\bibfnamefont {X.}~\bibnamefont {Shi}}, \bibinfo {author}
  {\bibfnamefont {A.}~\bibnamefont {van Roekeghem}}, \bibinfo {author}
  {\bibfnamefont {T.}~\bibnamefont {Qian}}, \bibinfo {author} {\bibfnamefont
  {E.}~\bibnamefont {Razzoli}}, \bibinfo {author} {\bibfnamefont
  {E.}~\bibnamefont {Rienks}}, \bibinfo {author} {\bibfnamefont {G.-F.}\
  \bibnamefont {Chen}}, \bibinfo {author} {\bibfnamefont {E.}~\bibnamefont
  {Ieki}}, \bibinfo {author} {\bibfnamefont {K.}~\bibnamefont {Nakayama}},
  \bibinfo {author} {\bibfnamefont {T.}~\bibnamefont {Sato}}, \bibinfo {author}
  {\bibfnamefont {T.}~\bibnamefont {Takahashi}}, \bibinfo {author}
  {\bibfnamefont {M.}~\bibnamefont {Shi}}, \ and\ \bibinfo {author}
  {\bibfnamefont {H.}~\bibnamefont {Ding}},\ }\href {\doibase
  10.1103/PhysRevB.88.220508} {\bibfield  {journal} {\bibinfo  {journal} {Phys.
  Rev. B}\ }\textbf {\bibinfo {volume} {88}},\ \bibinfo {pages} {220508}
  (\bibinfo {year} {2013})}\BibitemShut {NoStop}%
\bibitem [{\citenamefont {Hardy}\ \emph {et~al.}(2016)\citenamefont {Hardy},
  \citenamefont {B\"ohmer}, \citenamefont {de' Medici}, \citenamefont {Capone},
  \citenamefont {Giovannetti}, \citenamefont {Eder}, \citenamefont {Wang},
  \citenamefont {He}, \citenamefont {Wolf}, \citenamefont {Schweiss},
  \citenamefont {Heid}, \citenamefont {Herbig}, \citenamefont {Adelmann},
  \citenamefont {Fisher},\ and\ \citenamefont {Meingast}}]{Hardy:2016}%
  \BibitemOpen
  \bibfield  {author} {\bibinfo {author} {\bibfnamefont {F.}~\bibnamefont
  {Hardy}}, \bibinfo {author} {\bibfnamefont {A.~E.}\ \bibnamefont {B\"ohmer}},
  \bibinfo {author} {\bibfnamefont {L.}~\bibnamefont {de' Medici}}, \bibinfo
  {author} {\bibfnamefont {M.}~\bibnamefont {Capone}}, \bibinfo {author}
  {\bibfnamefont {G.}~\bibnamefont {Giovannetti}}, \bibinfo {author}
  {\bibfnamefont {R.}~\bibnamefont {Eder}}, \bibinfo {author} {\bibfnamefont
  {L.}~\bibnamefont {Wang}}, \bibinfo {author} {\bibfnamefont {M.}~\bibnamefont
  {He}}, \bibinfo {author} {\bibfnamefont {T.}~\bibnamefont {Wolf}}, \bibinfo
  {author} {\bibfnamefont {P.}~\bibnamefont {Schweiss}}, \bibinfo {author}
  {\bibfnamefont {R.}~\bibnamefont {Heid}}, \bibinfo {author} {\bibfnamefont
  {A.}~\bibnamefont {Herbig}}, \bibinfo {author} {\bibfnamefont
  {P.}~\bibnamefont {Adelmann}}, \bibinfo {author} {\bibfnamefont {R.~A.}\
  \bibnamefont {Fisher}}, \ and\ \bibinfo {author} {\bibfnamefont
  {C.}~\bibnamefont {Meingast}},\ }\href {\doibase 10.1103/PhysRevB.94.205113}
  {\bibfield  {journal} {\bibinfo  {journal} {Phys. Rev. B}\ }\textbf {\bibinfo
  {volume} {94}},\ \bibinfo {pages} {205113} (\bibinfo {year}
  {2016})}\BibitemShut {NoStop}%
\bibitem [{\citenamefont {Ikeda}\ \emph {et~al.}(2010)\citenamefont {Ikeda},
  \citenamefont {Arita},\ and\ \citenamefont {Kune\ifmmode~\check{s}\else
  \v{s}\fi{}}}]{Ikeda:2010}%
  \BibitemOpen
  \bibfield  {author} {\bibinfo {author} {\bibfnamefont {H.}~\bibnamefont
  {Ikeda}}, \bibinfo {author} {\bibfnamefont {R.}~\bibnamefont {Arita}}, \ and\
  \bibinfo {author} {\bibfnamefont {J.}~\bibnamefont
  {Kune\ifmmode~\check{s}\else \v{s}\fi{}}},\ }\href {\doibase
  10.1103/PhysRevB.81.054502} {\bibfield  {journal} {\bibinfo  {journal} {Phys.
  Rev. B}\ }\textbf {\bibinfo {volume} {81}},\ \bibinfo {pages} {054502}
  (\bibinfo {year} {2010})}\BibitemShut {NoStop}%
\bibitem [{\citenamefont {Kuroki}\ \emph {et~al.}(2009)\citenamefont {Kuroki},
  \citenamefont {Usui}, \citenamefont {Onari}, \citenamefont {Arita},\ and\
  \citenamefont {Aoki}}]{Kuroki:2009a}%
  \BibitemOpen
  \bibfield  {author} {\bibinfo {author} {\bibfnamefont {K.}~\bibnamefont
  {Kuroki}}, \bibinfo {author} {\bibfnamefont {H.}~\bibnamefont {Usui}},
  \bibinfo {author} {\bibfnamefont {S.}~\bibnamefont {Onari}}, \bibinfo
  {author} {\bibfnamefont {R.}~\bibnamefont {Arita}}, \ and\ \bibinfo {author}
  {\bibfnamefont {H.}~\bibnamefont {Aoki}},\ }\href {\doibase
  10.1103/PhysRevB.79.224511} {\bibfield  {journal} {\bibinfo  {journal} {Phys.
  Rev. B}\ }\textbf {\bibinfo {volume} {79}},\ \bibinfo {pages} {224511}
  (\bibinfo {year} {2009})}\BibitemShut {NoStop}%
\bibitem [{\citenamefont {Metzner}\ \emph {et~al.}(2012)\citenamefont
  {Metzner}, \citenamefont {Salmhofer}, \citenamefont {Honerkamp},
  \citenamefont {Meden},\ and\ \citenamefont {Sch\"onhammer}}]{Metzner:2012}%
  \BibitemOpen
  \bibfield  {author} {\bibinfo {author} {\bibfnamefont {W.}~\bibnamefont
  {Metzner}}, \bibinfo {author} {\bibfnamefont {M.}~\bibnamefont {Salmhofer}},
  \bibinfo {author} {\bibfnamefont {C.}~\bibnamefont {Honerkamp}}, \bibinfo
  {author} {\bibfnamefont {V.}~\bibnamefont {Meden}}, \ and\ \bibinfo {author}
  {\bibfnamefont {K.}~\bibnamefont {Sch\"onhammer}},\ }\href {\doibase
  10.1103/RevModPhys.84.299} {\bibfield  {journal} {\bibinfo  {journal} {Rev.
  Mod. Phys.}\ }\textbf {\bibinfo {volume} {84}},\ \bibinfo {pages} {299}
  (\bibinfo {year} {2012})}\BibitemShut {NoStop}%
\bibitem [{\citenamefont {Platt}\ \emph {et~al.}(2014)\citenamefont {Platt},
  \citenamefont {Hanke},\ and\ \citenamefont {Thomale}}]{Platt:2014}%
  \BibitemOpen
  \bibfield  {author} {\bibinfo {author} {\bibfnamefont {C.}~\bibnamefont
  {Platt}}, \bibinfo {author} {\bibfnamefont {W.}~\bibnamefont {Hanke}}, \ and\
  \bibinfo {author} {\bibfnamefont {R.}~\bibnamefont {Thomale}},\ }\href
  {\doibase 10.1080/00018732.2013.862020} {\bibfield  {journal} {\bibinfo
  {journal} {Adv. Phys.}\ }\textbf {\bibinfo {volume} {62}},\ \bibinfo {pages}
  {453} (\bibinfo {year} {2014})}\BibitemShut {NoStop}%
\bibitem [{\citenamefont {Graser}\ \emph
  {et~al.}(2010{\natexlab{a}})\citenamefont {Graser}, \citenamefont
  {Hirschfeld}, \citenamefont {Kopp}, \citenamefont {Gutser}, \citenamefont
  {Andersen},\ and\ \citenamefont {Mannhart}}]{Graser:2010}%
  \BibitemOpen
  \bibfield  {author} {\bibinfo {author} {\bibfnamefont {S.}~\bibnamefont
  {Graser}}, \bibinfo {author} {\bibfnamefont {P.~J.}\ \bibnamefont
  {Hirschfeld}}, \bibinfo {author} {\bibfnamefont {T.}~\bibnamefont {Kopp}},
  \bibinfo {author} {\bibfnamefont {R.}~\bibnamefont {Gutser}}, \bibinfo
  {author} {\bibfnamefont {B.~M.}\ \bibnamefont {Andersen}}, \ and\ \bibinfo
  {author} {\bibfnamefont {J.}~\bibnamefont {Mannhart}},\ }\href {\doibase
  doi:10.1038/nphys1687} {\bibfield  {journal} {\bibinfo  {journal} {Nature
  Phys.}\ }\textbf {\bibinfo {volume} {6}},\ \bibinfo {pages} {609} (\bibinfo
  {year} {2010}{\natexlab{a}})}\BibitemShut {NoStop}%
\bibitem [{\citenamefont {Graser}\ \emph
  {et~al.}(2010{\natexlab{b}})\citenamefont {Graser}, \citenamefont {Kemper},
  \citenamefont {Maier}, \citenamefont {Cheng}, \citenamefont {Hirschfeld},\
  and\ \citenamefont {Scalapino}}]{Graser:2010b}%
  \BibitemOpen
  \bibfield  {author} {\bibinfo {author} {\bibfnamefont {S.}~\bibnamefont
  {Graser}}, \bibinfo {author} {\bibfnamefont {A.~F.}\ \bibnamefont {Kemper}},
  \bibinfo {author} {\bibfnamefont {T.~A.}\ \bibnamefont {Maier}}, \bibinfo
  {author} {\bibfnamefont {H.-P.}\ \bibnamefont {Cheng}}, \bibinfo {author}
  {\bibfnamefont {P.~J.}\ \bibnamefont {Hirschfeld}}, \ and\ \bibinfo {author}
  {\bibfnamefont {D.~J.}\ \bibnamefont {Scalapino}},\ }\href {\doibase
  10.1103/PhysRevB.81.214503} {\bibfield  {journal} {\bibinfo  {journal} {Phys.
  Rev. B}\ }\textbf {\bibinfo {volume} {81}},\ \bibinfo {pages} {214503}
  (\bibinfo {year} {2010}{\natexlab{b}})}\BibitemShut {NoStop}%
\bibitem [{\citenamefont {Miyake}\ \emph {et~al.}(2010)\citenamefont {Miyake},
  \citenamefont {Nakamura}, \citenamefont {Arita},\ and\ \citenamefont
  {Imada}}]{Miyake:2010}%
  \BibitemOpen
  \bibfield  {author} {\bibinfo {author} {\bibfnamefont {T.}~\bibnamefont
  {Miyake}}, \bibinfo {author} {\bibfnamefont {K.}~\bibnamefont {Nakamura}},
  \bibinfo {author} {\bibfnamefont {R.}~\bibnamefont {Arita}}, \ and\ \bibinfo
  {author} {\bibfnamefont {M.}~\bibnamefont {Imada}},\ }\href {\doibase
  10.1143/JPSJ.79.044705} {\bibfield  {journal} {\bibinfo  {journal} {J. Phys.
  Soc. Japan}\ }\textbf {\bibinfo {volume} {79}},\ \bibinfo {pages} {044705}
  (\bibinfo {year} {2010})}\BibitemShut {NoStop}%
\bibitem [{\citenamefont {Platt}\ \emph {et~al.}(2012)\citenamefont {Platt},
  \citenamefont {Thomale}, \citenamefont {Honerkamp}, \citenamefont {Zhang},\
  and\ \citenamefont {Hanke}}]{Platt:2012}%
  \BibitemOpen
  \bibfield  {author} {\bibinfo {author} {\bibfnamefont {C.}~\bibnamefont
  {Platt}}, \bibinfo {author} {\bibfnamefont {R.}~\bibnamefont {Thomale}},
  \bibinfo {author} {\bibfnamefont {C.}~\bibnamefont {Honerkamp}}, \bibinfo
  {author} {\bibfnamefont {S.-C.}\ \bibnamefont {Zhang}}, \ and\ \bibinfo
  {author} {\bibfnamefont {W.}~\bibnamefont {Hanke}},\ }\href {\doibase
  10.1103/PhysRevB.85.180502} {\bibfield  {journal} {\bibinfo  {journal} {Phys.
  Rev. B}\ }\textbf {\bibinfo {volume} {85}},\ \bibinfo {pages} {180502}
  (\bibinfo {year} {2012})}\BibitemShut {NoStop}%
\bibitem [{\citenamefont {Wang}\ \emph {et~al.}(2014)\citenamefont {Wang},
  \citenamefont {Eberlein},\ and\ \citenamefont {Metzner}}]{WangJ:2014}%
  \BibitemOpen
  \bibfield  {author} {\bibinfo {author} {\bibfnamefont {J.}~\bibnamefont
  {Wang}}, \bibinfo {author} {\bibfnamefont {A.}~\bibnamefont {Eberlein}}, \
  and\ \bibinfo {author} {\bibfnamefont {W.}~\bibnamefont {Metzner}},\ }\href
  {\doibase 10.1103/PhysRevB.89.121116} {\bibfield  {journal} {\bibinfo
  {journal} {Phys. Rev. B}\ }\textbf {\bibinfo {volume} {89}},\ \bibinfo
  {pages} {121116} (\bibinfo {year} {2014})}\BibitemShut {NoStop}%
\bibitem [{\citenamefont {Kemper}\ \emph {et~al.}(2010)\citenamefont {Kemper},
  \citenamefont {Maier}, \citenamefont {Graser}, \citenamefont {Cheng},
  \citenamefont {Hirschfeld},\ and\ \citenamefont {Scalapino}}]{kemper:2010}%
  \BibitemOpen
  \bibfield  {author} {\bibinfo {author} {\bibfnamefont {A.~F.}\ \bibnamefont
  {Kemper}}, \bibinfo {author} {\bibfnamefont {T.~A.}\ \bibnamefont {Maier}},
  \bibinfo {author} {\bibfnamefont {S.}~\bibnamefont {Graser}}, \bibinfo
  {author} {\bibfnamefont {H.-P.}\ \bibnamefont {Cheng}}, \bibinfo {author}
  {\bibfnamefont {P.~J.}\ \bibnamefont {Hirschfeld}}, \ and\ \bibinfo {author}
  {\bibfnamefont {D.~J.}\ \bibnamefont {Scalapino}},\ }\href {\doibase
  10.1088/1367-2630/12/7/073030} {\bibfield  {journal} {\bibinfo  {journal}
  {New J. Phys.}\ }\textbf {\bibinfo {volume} {12}},\ \bibinfo {pages} {073030}
  (\bibinfo {year} {2010})}\BibitemShut {NoStop}%
\bibitem [{\citenamefont {Raphael}\ \emph {et~al.}(1998)\citenamefont
  {Raphael}, \citenamefont {Reeves},\ and\ \citenamefont
  {Skelton}}]{Raphael:1998}%
  \BibitemOpen
  \bibfield  {author} {\bibinfo {author} {\bibfnamefont {M.~P.}\ \bibnamefont
  {Raphael}}, \bibinfo {author} {\bibfnamefont {M.~E.}\ \bibnamefont {Reeves}},
  \ and\ \bibinfo {author} {\bibfnamefont {E.~F.}\ \bibnamefont {Skelton}},\
  }\href {\doibase 10.1063/1.1148780} {\bibfield  {journal} {\bibinfo
  {journal} {Rev. Sci. Instrum.}\ }\textbf {\bibinfo {volume} {69}},\ \bibinfo
  {pages} {1451} (\bibinfo {year} {1998})}\BibitemShut {NoStop}%
\bibitem [{\citenamefont {Shatz}\ \emph {et~al.}(1993)\citenamefont {Shatz},
  \citenamefont {Shaulov},\ and\ \citenamefont {Yeshurun}}]{Shatz:1993}%
  \BibitemOpen
  \bibfield  {author} {\bibinfo {author} {\bibfnamefont {S.}~\bibnamefont
  {Shatz}}, \bibinfo {author} {\bibfnamefont {A.}~\bibnamefont {Shaulov}}, \
  and\ \bibinfo {author} {\bibfnamefont {Y.}~\bibnamefont {Yeshurun}},\ }\href
  {\doibase 10.1103/PhysRevB.48.13871} {\bibfield  {journal} {\bibinfo
  {journal} {Phys. Rev. B}\ }\textbf {\bibinfo {volume} {48}},\ \bibinfo
  {pages} {13871} (\bibinfo {year} {1993})}\BibitemShut {NoStop}%
\bibitem [{\citenamefont {Mazin}\ \emph {et~al.}(2010)\citenamefont {Mazin},
  \citenamefont {Devereaux}, \citenamefont {Analytis}, \citenamefont {Chu},
  \citenamefont {Fisher}, \citenamefont {Muschler},\ and\ \citenamefont
  {Hackl}}]{Mazin:2010a}%
  \BibitemOpen
  \bibfield  {author} {\bibinfo {author} {\bibfnamefont {I.~I.}\ \bibnamefont
  {Mazin}}, \bibinfo {author} {\bibfnamefont {T.~P.}\ \bibnamefont
  {Devereaux}}, \bibinfo {author} {\bibfnamefont {J.~G.}\ \bibnamefont
  {Analytis}}, \bibinfo {author} {\bibfnamefont {J.-H.}\ \bibnamefont {Chu}},
  \bibinfo {author} {\bibfnamefont {I.~R.}\ \bibnamefont {Fisher}}, \bibinfo
  {author} {\bibfnamefont {B.}~\bibnamefont {Muschler}}, \ and\ \bibinfo
  {author} {\bibfnamefont {R.}~\bibnamefont {Hackl}},\ }\href {\doibase
  10.1103/PhysRevB.82.180502} {\bibfield  {journal} {\bibinfo  {journal} {Phys.
  Rev. B}\ }\textbf {\bibinfo {volume} {82}},\ \bibinfo {pages} {180502}
  (\bibinfo {year} {2010})}\BibitemShut {NoStop}%
\bibitem [{B1g()}]{B1g}%
  \BibitemOpen
  \href@noop {} {}\bibinfo {note} {Note that the Fe $B_{1g}$ phonon of the
  crystallographic (2\,Fe) unit cell appears in $B_{2g}$ in the 1\,Fe unit cell
  used here for the analysis of the electronic spectra.}\BibitemShut {Stop}%
\bibitem [{\citenamefont {Pekker}\ and\ \citenamefont
  {Varma}(2015)}]{Pekker:2015}%
  \BibitemOpen
  \bibfield  {author} {\bibinfo {author} {\bibfnamefont {D.}~\bibnamefont
  {Pekker}}\ and\ \bibinfo {author} {\bibfnamefont {C.}~\bibnamefont {Varma}},\
  }\href {\doibase 10.1146/annurev-conmatphys-031214-014350} {\bibfield
  {journal} {\bibinfo  {journal} {Ann. Rev. Cond. Mat. Phys.}\ }\textbf
  {\bibinfo {volume} {6}},\ \bibinfo {pages} {269} (\bibinfo {year}
  {2015})}\BibitemShut {NoStop}%
\end{thebibliography}%
\clearpage

\begin{appendix}
\label{sec:appendix}

\section{Theoretical Methods}
\label{apptheo}
\subsection{Model Hamiltonian}
To determine the relative strength of $s$- and $d$-wave pairing channels in Ba$_{1-x}$K$_x$Fe$_2$As$_2$ from a model Hamiltonian, the results of which are depicted in Fig.~3\,(b) and (c), we start with the five iron $d$-orbital approach proposed by Graser \textit{et al}.~\cite{Graser:2010} for parent BaFe$_2$As$_2$. The tight-binding model,
\begin{equation}
  \mathcal{H}_0 = \sum_{\mathbf{k},s}\sum_{a,b=1}^5 c^{\dagger}_{\mathbf{k}as}K_{ab}(\mathbf{k})c^{\phantom{\dagger}}_{\mathbf{k}bs},
\label{eq:H0}
\end{equation}
with  $\mathbf{k}$, $a$, $s$ denoting momentum, orbital and spin degrees of freedom, accurately reproduces the \textit{ab initio} band structure of BaFe$_2$As$_2$ at low energies \cite{Graser:2010b}. The different doping levels are modeled by a rigid band shift, which amounts to $0.1$\,eV or roughly $2.5\%$ of the bandwidth for $x=0.5$ potassium substitution. Assuming a pronounced two-dimensional character of the Fermi surface sheets, we restrict the pairing calculation to the 2D cut of the Fermi surface for $k_z=0$.

For the interaction part $\mathcal{H}_{\rm int}$, we use a complete set of on-site intra- and interorbital repulsion as well as Hund's and pair-hopping terms,
\begin{align}
\label{eq:H-int}
\mathcal{H}_{\rm int} = \sum_i &\left[U_{\rm intra} \sum_a n_{ia\uparrow}n_{ia\downarrow} + U_{\rm inter} \sum_{a< b,ss^\prime} n_{ias}n_{ibs^\prime}\right. \\\nonumber
&\left. + J_H\sum_{a< b} \vec{S}_{ia}\vec{S}_{ib} + J_{\rm pair}  \sum_{a<b}c^{\dagger}_{ia\uparrow}c^{\dagger}_{ia\downarrow}
c^{\phantom{\dagger}}_{ib\uparrow}c^{\phantom{\dagger}}_{ib\downarrow}\right].
\end{align}
Both fRG and RPA are independently applied to this microsopic model. The parameter set chosen in fRG is $U_{\rm intra} = 4.0$\,eV, $U_{\rm inter} = 2.0$\,eV and $J_H = J_{\rm pair} = 0.7$\,eV which are close to constrained RPA estimations \cite{Miyake:2010}. In order to avoid a magnetic instability in RPA, we choose interaction parameters that are smaller than those used in the fRG calculation. We set $U=0.85$\,eV and $U'=U/2$, $J=J'=U/4$. For these parameters, we find that the pairing interaction vertex $V(k,q)$ is determined by spin fluctuations, and orbital fluctuations do not play a significant role. In the fRG, the bare interactions are renormalized to smaller values due to the coupling between the different particle-hole and particle-particle channels. This coupling is absent in the RPA, which is consistent with the need to use smaller bare parameters for the latter.

\subsection{fRG}
The method of fRG proved very successful for the prediction and exploration of the interplay between magnetic and superconducting order in iron-based compounds \cite{Platt:2014,Thomale:2011a,Platt:2012}. Its main strength lies in its ability to treat different ordering channels on an equal footing, whilst simultaneously  allowing for the study of microscopic models with multiple degrees of freedom. Similar to the familiar fRG concept of thinning out degrees  down to a low-energy cutoff $\Lambda$ which serves as an upper bound for the transition temperature while leaving the low-energy physics invariant \cite{Metzner:2012,Platt:2014}, the fRG method provides an effective low-energy theory $\mathcal{H}^\Lambda$ which immediately (or at least more clearly) reveals the favored ordering tendencies. In this flow to low energies, different fluctuations such as spin and charge density wave, Pomeranchuk, and superconducting ordering tendencies are included on equal footing \cite{Thomale:2011a}.

Without going into the technical details of the calculations, it is important to mention that one usually implements the fRG in terms of an appropriate band basis $\gamma_k^\dagger$, $\gamma_k$ which diagonalizes the quadratic part Eq.~(\ref{eq:H0}). The effective low-energy description in this basis then reads
\begin{equation}
\label{eq:HLambda}
\mathcal{H}^\Lambda = \mathcal{H}_0^\Lambda + \sum \limits_{k_1,...k_4} \tilde{V}^\Lambda(k_1,k_2,k_3,k_4) \gamma_{k_1}^\dagger \gamma_{k_2}^\dagger \gamma_{k_3} \gamma_{k_4}\,,
\end{equation}
with a renormalized quadratic part $\mathcal{H}_0^\Lambda$ and an effective two-particle interaction $\tilde{V}^\Lambda$. In the fRG flow, spontaneous symmetry breaking is signalled by diverging effective interactions at a critical energy scale $\Lambda_c$. In the``standard'' fRG, the 2-particle vertex $\tilde{V}^\Lambda$ is computed from an fRG flow (integrated) down to a ``mean-field'' scale $\Lambda_\mathrm{MF}$ slightly above the critical scale $\Lambda_c$. This first part of our calculation is similar to our earlier study of hole-doped ${\rm Ba_{1-x}K_xFe_2As_2}$~\cite{Thomale:2011a}, where further details can be found.

However, to compare the fRG leading eigenvalues in the $s_\pm$- and $d$-channels with experiment and with the RPA results, based on the eigenvalue description of the dominant and subdominant Raman peaks, we have to ``roll back'' this particle-particle-channel to obtain the \textit{pp-irreducible pairing interaction}. This is discussed in Ref.~\onlinecite{WangJ:2014}  with application to the single band Hubbard model. It amounts to solving a linear integral (Bethe-Salpeter) equation for the pp-irreducible vertex $V(k,q)$, where $V(k,k') \equiv V(k,k',k',k)$, i.e.
\begin{equation}
\label{eq:PPIrreducible}
\tilde{V}^\Lambda(k,q)=V^\Lambda(k,q)-\int_p V^\Lambda(k,p) G^\Lambda(p) G^\Lambda(-p) \tilde{V}^\Lambda(p,q)\,,
\end{equation}
schematically shown in Fig.~\ref{sfig:PPIrreducible} below. Here $\tilde{V}^\Lambda(k,q)$ denotes the usual (1-particle irreducible) vertex obtained from the "standard" fRG, including all modes down to $\Lambda_\mathrm{MF}$, which is chosen slightly above the critical scale $\Lambda_c$: the explicit solution is obtained as $V^\Lambda = \tilde{V}^\Lambda (1-D^\Lambda \tilde{V}^\Lambda)^{-1}$, i.e. matrix inversion, due to  the usual discretization of momentum space in patches rendering the kernel of Eq.~(\ref{eq:PPIrreducible}) separable. This kernel, or diagonal matrix, $D$ is radially integrated in each patch with the frequency sum in Eq.~(\ref{eq:H0}) evaluated analytically.
\begin{figure}[htb]
  \centering
  \vspace{1mm}
  \includegraphics[scale=1]{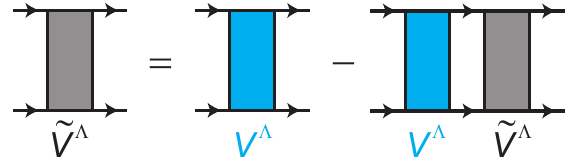}
  \caption{Diagrammatic relation between $V^\Lambda$ and $\tilde{V}^\Lambda$. Fermionic propagators are illustrated by solid black lines, the interactions $V^\Lambda$ and $\tilde{V}^\Lambda$ by blue and grey boxes, respectively.}
  \label{sfig:PPIrreducible}
\end{figure}

This pp-irreducible fRG vertex is inserted in Eq.~(1) to obtain the eigenvalue decomposition $\lambda_\alpha$, where $\lambda_2/\lambda_1$ as well as $\lambda_3/\lambda_1$ is then compared against RPA and the experimental finding as described in the main text. (The integral measure in Eq.~(1) becomes $\int_{\text{FS}} dq \equiv \sum_i \int_{\text{FS}_i}d^2q/2\pi v_{\text{F}}(q)$, where the index $i$ runs over all Fermi surface pockets and $v_{\text{F}}(q)$ denotes the Fermi velocity.)

Choosing the pp-irreducible vertex within the fRG approach also substantially reduces the quantitative influence of the fRG-cutoff $\Lambda_\mathrm{MF}$: by rolling back the pp-channel to $V$ (Fig.~\ref{sfig:PPIrreducible}), one takes out a steeply growing logarithm. This can be most easily seen if the kernel $D$ is assumed to be a constant $d$. In this case, $V^\Lambda$ and $\tilde{V}^\Lambda$ would have the eigenvalues related by $\lambda=\frac{\tilde{\lambda}}{1-d\tilde{\lambda}}$. The reduction of the cutoff influence, provided the particle-hole correlations, i.e. spin-fluctuations, die out at low enough scales, then holds, and hence the pp-irreducible pairing interaction has saturated at $\Lambda_\mathrm{MF}$.

For K substitution larger than $x\approx 0.25$, the pairing fluctuations with the effective (renormalized) two-particle irreducible interactions $V^\Lambda_{\text{fRG}}(k,-k,q,-q)\equiv V^\Lambda(k,q)$ [Eqs.~(\ref{eq:HLambda}) and (\ref{eq:PPIrreducible})] are dominant in the low-energy regime, which justifies us to constrain ourselves to the Cooper channel and hence superconducting Fermi surface instabilities.

\subsection{RPA}

In the RPA approximation, the pair structure which arises form a spin-fluctuation exchange interaction is determined from the scattering vertex
\begin{eqnarray}
	\Gamma_{ij}(\bk,\bk') &=& {\rm Re} \sum_{\ell_1\ell_2\ell_3\ell_4} a_{\nu_i}^{\ell_1,*}(\bk)a_{\nu_i}^{\ell_4,*}(-\bk) \\ &&\hspace{-0.5cm}\times\Gamma_{\ell_1\ell_2\ell_3\ell_4}(\bk,\bk',\omega=0)
	 \,a_{\nu_j}^{\ell_2}(\bk')a_{\nu_j}^{\ell_3}(-\bk')\,.\nonumber
\end{eqnarray}
Here the momenta $\bk$ and $\bk'$ are restricted to the Fermi surface $\bk \in {\cal C}_i$ and $\bk' \in {\cal C}_j$ where $i$ and $j$ label the different pockets, and $a^{\ell}_{\nu}(\bk)$ are the orbital components of the band eigenvectors. The particle-particle scattering vertex between orbitals $\ell_1$, $\ell_4$ and $\ell_2$, $\ell_3$ is given by
\begin{eqnarray}
	\Gamma_{\ell_1\ell_2\ell_3\ell_4}(\bk,\bk',\omega) &=& \left[ \frac{3}{2}\bar{U}^s\chi_s^{\rm RPA}(\bk-\bk',\omega)\bar{U}^s+\frac{1}{2}\bar{U}^s\right.\\
	 && - \left.\frac{1}{2}\bar{U}^c\chi_c^{\rm RPA}(\bk-\bk',\omega)\bar{U}^c+\frac{1}{2}\bar{U}^c\right]\,.\nonumber
\end{eqnarray}
The interaction matrices in orbital space for spin and charge channels, $\bar{U}^s$ and $\bar{U}^c$ respectively, contain linear combinations of the interaction parameters $U$, $U'$, $J$ and $J'$ (see e.g. Ref.~\onlinecite{kemper:2010}). The RPA spin- ($\chi^{\rm RPA}_s$) and charge ($\chi^{\rm RPA}_c$) susceptibilities have the usual form
\begin{eqnarray}
	\chi^{\rm RPA}_{c/s,\ell_1\ell_2\ell_3\ell_4}({\bf q}) = \left\{\chi^0({\bf q})[1\pm\bar{U}^{c/s}\chi^0({\bf q})]^{-1}\right\}_{\ell_1\ell_2\ell_3\ell_4}
\end{eqnarray}
with the bare susceptibility matrix
\begin{eqnarray}
 	\chi^0_{\ell_1\ell_2\ell_3\ell_4}({\bf q},\omega) &=&\\&&\hspace{-2cm} -\frac{1}{N}\sum_{\bk,\mu\nu} a^{\ell_4}_\nu(\bk)a^{\ell_2,*}_{\nu}(\bk)a^{\ell_1}_\mu(\bk+{\bf q})a^{\ell_3,*}_{\mu}(\bk+{\bf q})\nonumber\\
 	&\times& \frac{f(E_\nu(\bk))-f(E_\mu(\bk+{\bf q}))}{\omega+i0^++E_\nu(\bk)-E_\mu(\bk+{\bf q})}\nonumber\,.
\end{eqnarray}
Here, $f(E)$ is the Fermi function and $E_\nu(\bk)$ is the energy dispersion for band $\nu$. The symmetry function $g(\bk)$ of the pairing state can then be found by solving an eigenvalue problem of the form
\begin{equation}
\label{eqrpa}
	-\sum_j\oint_{C_j}\frac{d\bk'_\parallel}{2\pi v_F(\bk'_\parallel)}\Gamma_{ij}(\bk,\bk')g_\alpha(\bk') = \lambda_\alpha g_\alpha(\bk)\,,
\end{equation}
where $v_F(\bk)$ is the Fermi velocity. The eigenfunction $g_\alpha(\bk)$ with the largest eigenvalue $\lambda_\alpha$ gives the leading instability (ground state) of the system and sub-leading instabilities have smaller $\lambda_\alpha$. We have executed all steps within the RPA formalism once again to illustrate the precise form equivalence between Eq.~(\ref{eqrpa}) and  Eq.~(1) in the main text upon the identification $\int_{\text{FS}}\equiv \sum_j\oint_{C_j}\frac{d\bk'_\parallel}{2\pi v_F(\bk'_\parallel)}$ as well as $V(k,q) \equiv \Gamma_{ij}(\bk,\bk')$.

\section{Samples}
\begin{figure}[htb]
  \centering
  \vspace{1mm}
  \includegraphics[width=\columnwidth]{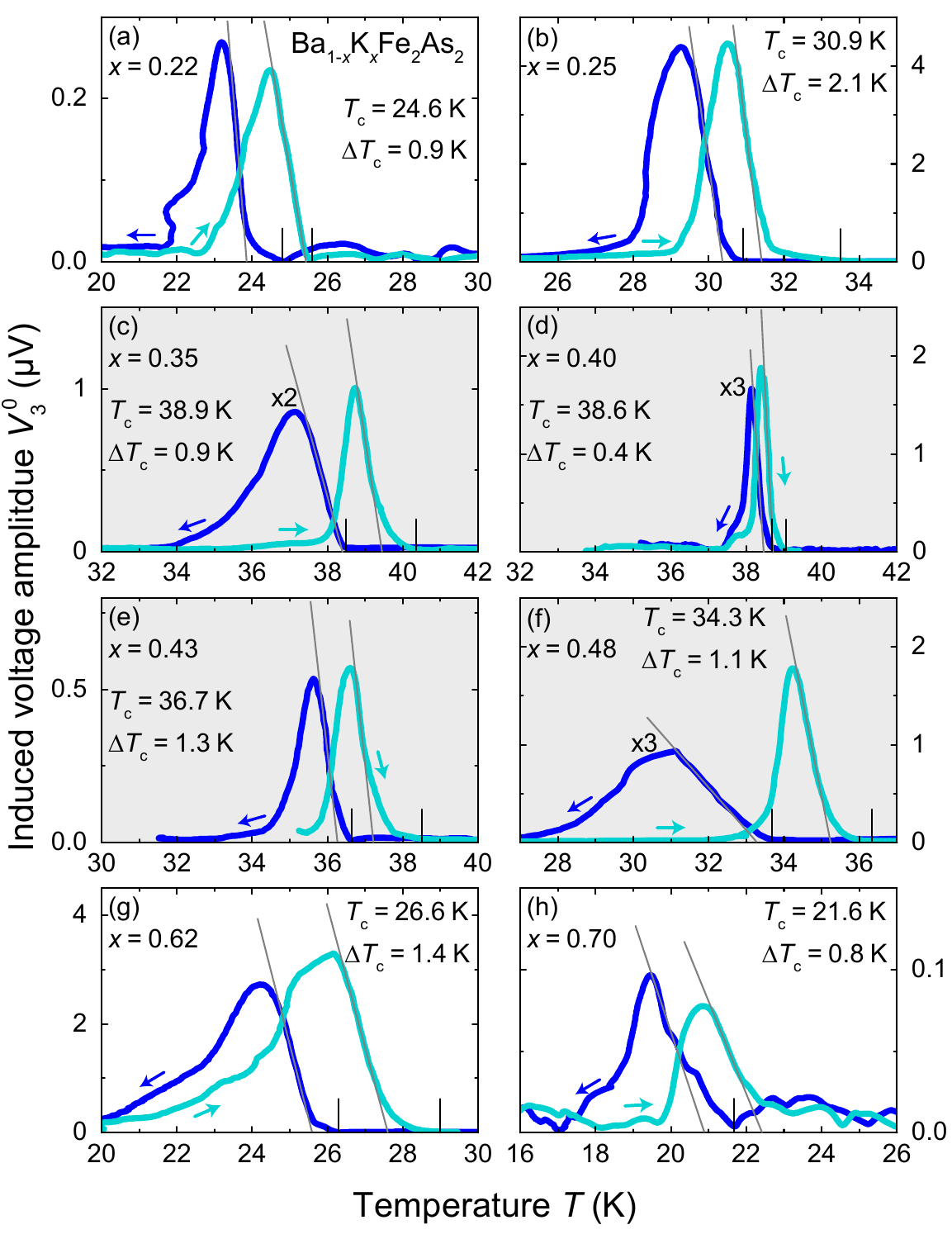}
  \caption{Non-linear susceptibility of all samples measured. The two data sets correspond to warm up (turquoise) and cool down (blue). $T_{c\pm}$ is the extrapolation of the linear part to zero voltage, the tail above $T_{c\pm}$ corresponds to the transition width $\Delta T_c$ \cite{Shatz:1993}.}
  \label{sfig:Tc}
\end{figure}

The single crystals of hole doped ${\rm Ba_{1-x}K_{x}Fe_2As_2}$ were grown using a self-flux technique and have been characterized elsewhere \cite{Shen:2011,Karkin:2014}. The potassium concentration was determined by microprobe analysis. For all samples studied we measured the non-linear magnetic susceptibility $\chi_m^{(3)}(T)$. The results are complied in Fig.~\ref{sfig:Tc}. $\chi_m^{(3)}(T)$ is particularly sensitive to inhomogeneities of the samples \cite{Raphael:1998}. The susceptibility was measured during a continuous warm up~(+)/cool down~(-) at a rate of typically $\pm2$\,K per minute. According to Shatz and coworkers \cite{Shatz:1993},  $T_{c\pm}$ is the extrapolation of the linear part to zero voltage at three times the excitation frequency $3f$ with $f=33$\,kHz. We define $T_c = 0.5(T_{c+} + T_{c-})$. The tail above $T_{c\pm}$ is a reliable estimate of the transition width $\Delta T_c$. For the Raman measurements samples with narrow superconducting transitions were selected having $\Delta T_c$ values in the range 0.4 to 2\,K. The doping levels and typical sample temperatures are displayed in Fig.~1\,(a). The parameters are compiled in Table~\ref{tab:Tc}.

\begin{table}[htb]
  \caption{Parameters of the $\mathrm{Ba_{1-x}K_{x}Fe_2As_2}$ samples studied. The doping levels of the samples used for the quantitative analysis are shown in bold face. Samples labeled with a and b were prepared in the laboratories of H.-H. Wen \cite{Shen:2011} and T. Wolf \cite{Karkin:2014}, respectively.}
  \vspace{1mm}
  \centering
  \begin{tabular}{c c c c}
  \hline\hline\\[-2ex]
  ~K doping $x$~& ~~~~$T_c$\,(K)~~~~ & ~~~$\Delta T_c$\,(K)~~~ & ~~source~~\\
  \hline\\[-2ex]
  $0.22$      &24.6&0.9&b\\
  $0.25$      &30.9&2.1&a\\
  $\bf{0.35}$ &38.9&0.9&b\\
  $\bf{0.40}$ &38.6&0.4&a\\
  $\bf{0.43}$ &36.7&1.3&b\\
  $\bf{0.48}$ &34.3&1.1&b\\
  $0.62$      &26.6&1.4&b\\
  $0.70$      &21.6&0.8&b\\
  \hline
  \hline
  \end{tabular}
  \label{tab:Tc}
\end{table}

\section{Experiment}
The experiments were performed with standard light scattering equipment. For excitation a solid state laser (Coherent, Genesis MX\,SLM) was used emitting at 575\,nm. The samples were mounted on the cold finger of a He-flow cryostat in a cryogenically pumped vacuum. The laser-induced heating was determined experimentally to be close to 1\,K per mW absorbed laser power. Spectra were measured in the four polarization configurations $xy$, $x^\prime y^\prime$, $RR$, and $RL$ where $x$ and $y$ are along the Fe-Fe bonds, $x^\prime = 1/\sqrt{2}(x+y)$, $y^\prime = 1/\sqrt{2}(y-x)$, and $R/L = 1/\sqrt{2}(x\pm iy)$. All symmetry components ($A_{1g}$, $A_{2g}$, $B_{1g}$, and  $B_{2g}$ for tetragonal ${\rm Ba_{1-x}K_{x}Fe_2As_2}$) can be extracted using linear combinations of the experimental spectra. For the symmetry assignment we use the 1\,Fe per unit cell [cf. Fig.~1\,(b) for the corresponding BZ] \cite{Muschler:2009,Mazin:2010a}. The spectra we show within this work represent the response $R\chi^{\prime\prime}(\Omega,T)$ which is obtained by dividing the cross section by the Bose-Einstein factor $\{1+n(T,\Omega)\}=[1-\exp(-\hbar\Omega/k_BT)]^{-1}$ in which $R$ is an experimental constant. In some cases we isolate superconductivity-induced contributions by subtracting the response measured at $T\gtrsim T_c$ from the spectra taken at $T\ll T_c$ and label the difference spectra $\Delta R\chi^{\prime\prime}(\Omega,T)$.

\section{Resonance effects}
\label{Asec:Resonance}

\begin{figure}[htb]
  \centering
  \vspace{1mm}
  \includegraphics[scale=1]{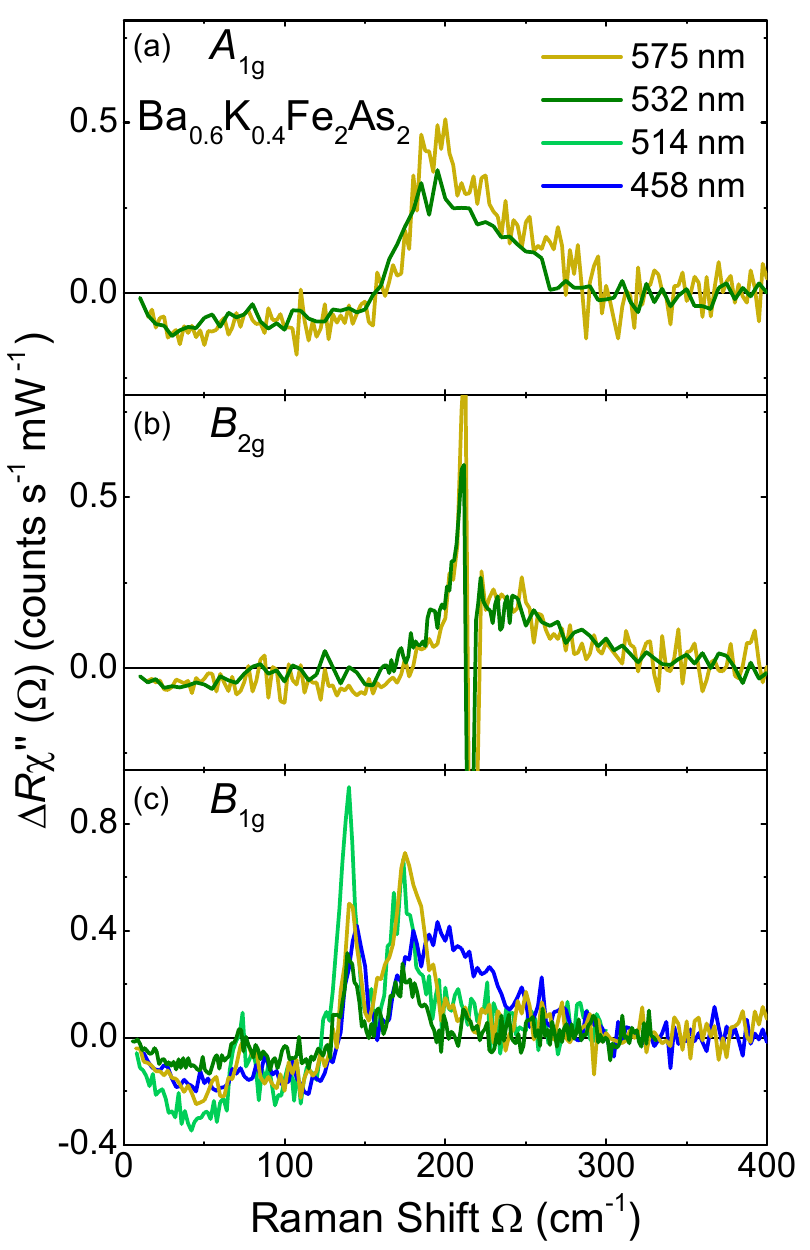}
  \caption{Resonance effects in $\mathrm{Ba_{1-x}K_{x}Fe_2As_2}$ at $x=0.40$. Shown are the difference spectra $\Delta R\chi^{\prime\prime}(\Omega) = R\chi^{\prime\prime}(\Omega,T=10\,{\rm K}) - R\chi^{\prime\prime}(\Omega,T\gtrsim T_c)$ for the three Raman-active symmetries (see text and  Fig.~2). The spectra are measured with the laser lines at 575, 532, and 458\,nm (only $B_{1g}$).
  }
  \label{sfig:Resonance}
\end{figure}

Figure~\ref{sfig:Resonance} shows the difference spectra $\Delta R\chi^{\prime\prime}(\Omega) = R\chi^{\prime\prime}(\Omega,T=10\,{\rm K}) - R\chi^{\prime\prime}(\Omega,T\gtrsim T_c)$ of the Raman reponse for excitation lines at 458, 514, 532, and 575\,nm. For non-resonant scattering processes one would expect the response to be independent of the excitation. This worked well for optimally doped Ba(Fe$_{0.939}$Co$_{0.061}$)$_2$As$_2$ \cite{Mazin:2010a}. Here, we observe little changes upon switching between green and yellow excitation in $A_{1g}$ and $B_{2g}$ symmetry [Fig.~\ref{sfig:Resonance}\,(a) and (b)]. In $B_{1g}$ symmetry [Fig.~\ref{sfig:Resonance}\,(c)], the overall intensity reduces for 532 \,nm with respect to 575 \,nm excitation wave length, whereas a substantial enhancement of spectral weight is found for 514 and 458\,nm. In the latter case both the pair-breaking peak and the BS mode shift. At first glance, this indicates a dependence on the details of the experiment. However, the BS mode does not change shape and by and large follows the intensity of the pair-breaking peaks. The change of the pair breaking effect for 458\,nm indicates the appearance of an orbital-dependent resonance. In the blue the enhancement occurs on those parts of the Fermi surface which have a larger gap. As a consequence, the collective mode which is directly derived from the pair breaking intensity shifts accordingly. Qualitatively, this is further support for the collective character of the mode at 140\,cm$^{-1}$ although the details need to be worked out in a future systematic study. Resonance effects must be considered for blue photons, but are weak for yellow photons and hence do not interfere with the analysis of this work.

\section{Symmetry-resolved spectra}
\label{Asec:SR}

\begin{figure}[htb]
  \centering
  \includegraphics[scale=1]{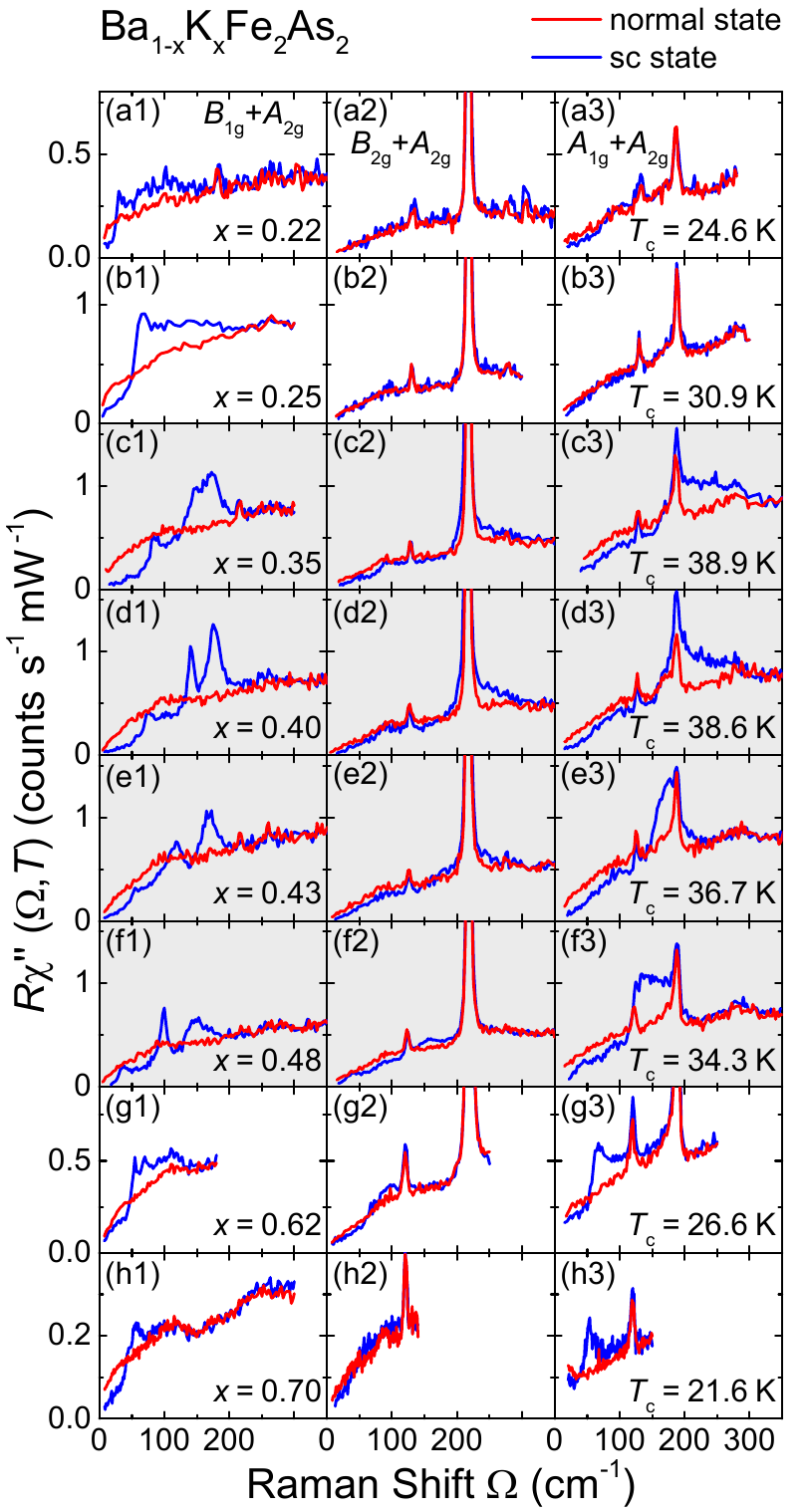}
  \caption{Doping-dependent Raman spectra. The Raman response $R\chi^{\prime\prime}(\Omega,T,x)$ (raw data after division by the Bose-Einstein factor) of ${\rm Ba_{1-x}K_{x}Fe_2As_2}$ is shown in $x^\prime y^\prime$ ($B_{1g}+A_{2g}$), $xy$ ($B_{2g}+A_{2g}$), and $RR$ ($A_{1g}+A_{2g}$) polarization above (red) and below (blue) $T_c$. The results around optimal doping, $0.35 \leq x \leq 0.48$, are highlighted.
  }
  \label{sfig:pol-all}
\end{figure}

Figure~\ref{sfig:pol-all} shows the raw data in the three main polarization configurations $xx$, $x^\prime y^\prime$ and $RR$ measured for this study at temperatures indicated in Fig.~1. The $x^\prime y^\prime$ normal-state spectra at $x=0.35$ are multiplied by 0.84 to make them match the superconducting spectra. The superposed narrow lines in $B_{2g}$ and $A_{1g}$ symmetry are Raman-active phonons \cite{B1g}.

\begin{figure}[htb]
  \centering
  \includegraphics[scale=1]{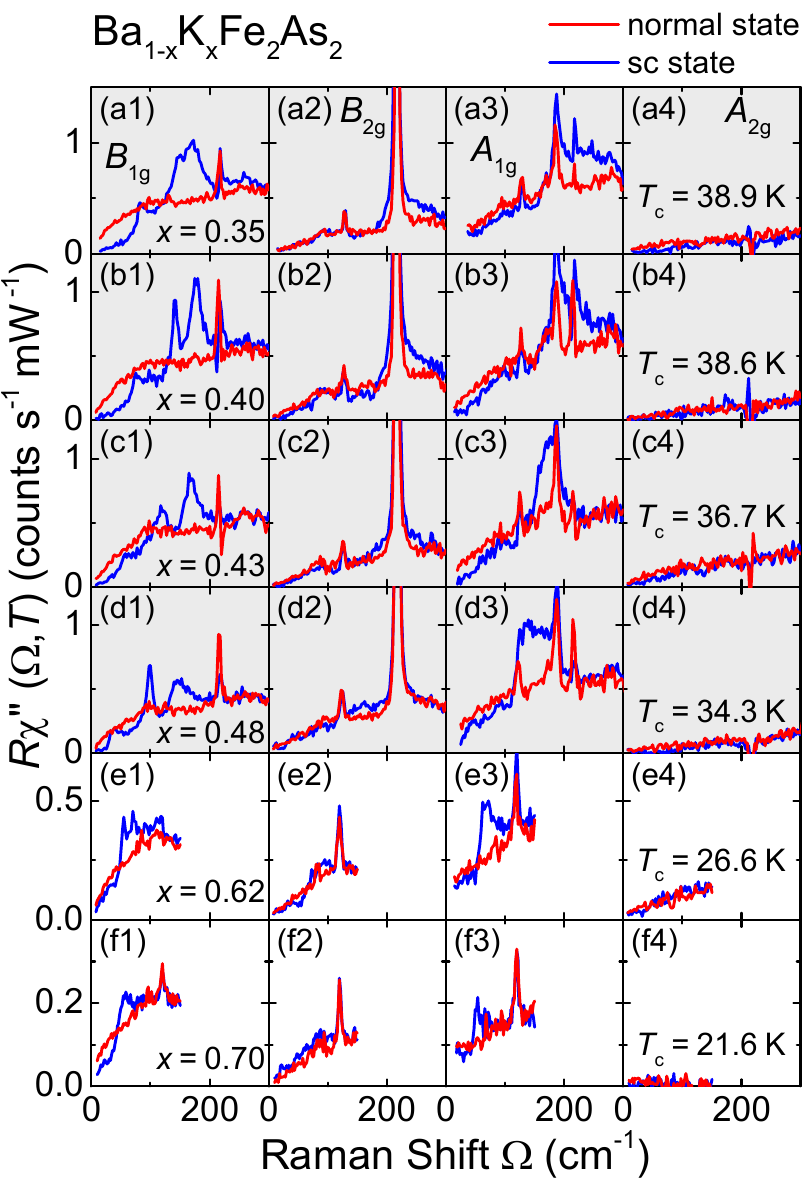}
  \caption{Symmetry-resolved spectra of $\mathrm{Ba_{1-x}K_{x}Fe_2As_2}$. The spectra for doping levels $x$ as indicated are measured at approximately 8\,K (blue) and above (red) the respective transition temperatures. The contributions from $A_{2g}$ symmetry are structureless, temperature independent, and substantially smaller than those in the other symmetries.
  }
  \label{sfig:SR}
\end{figure}

For a few samples we also measured spectra at intermediate temperatures which are partially published \cite{Bohm:2014}. For deriving the symmetry-resolved spectra also the $RL$ configuration is needed. The full symmetry analysis is available for $0.35\le x \le 0.70$ as shown in Fig.~\ref{sfig:SR}. The $A_{2g}$ spectra are temperature independent and typically on order 10\,\% of those of the other symmetries. They vanish for $x=0.7$. 
Therefore, for avoiding subtraction procedures, reducing the time for the measurements and improving the statistics we can use the spectra measured at $xx$, $x^\prime y^\prime$ and $RR$ (Fig.~\ref{sfig:pol-all}).

Except for the spectra at $x=0.22$ and 0.25 there are pair-breaking features in all symmetries (except for $A_{2g}$). In general, the structures are narrower and more pronounced in $B_{1g}$ than in $A_{1g}$ and $B_{2g}$ symmetry (see also Fig.~\ref{sfig:SR}). The $B_{1g}$ spectra at $x=0.22$, 0.25, 0.62, and 0.70 [Fig.~\ref{sfig:pol-all}\,(a1), (b1), (g1), (h1)] are qualitatively different from those around optimal doping [Fig.~\ref{sfig:pol-all}\,(c1), (d1), (e1), (f1)]. Whereas a narrow mode or an indication thereof is still found along with an isolated broad maximum for $x=0.62$ and $x=0.70$, respectively, there are only  broad maxima at 30\,cm$^{-1}$ and 65\,cm$^{-1}$ next to shoulders around 100 and 150\,cm$^{-1}$ for $x=0.22$ and 0.25, respectively.

\begin{figure}[htb]
  \centering
  \includegraphics[scale=1]{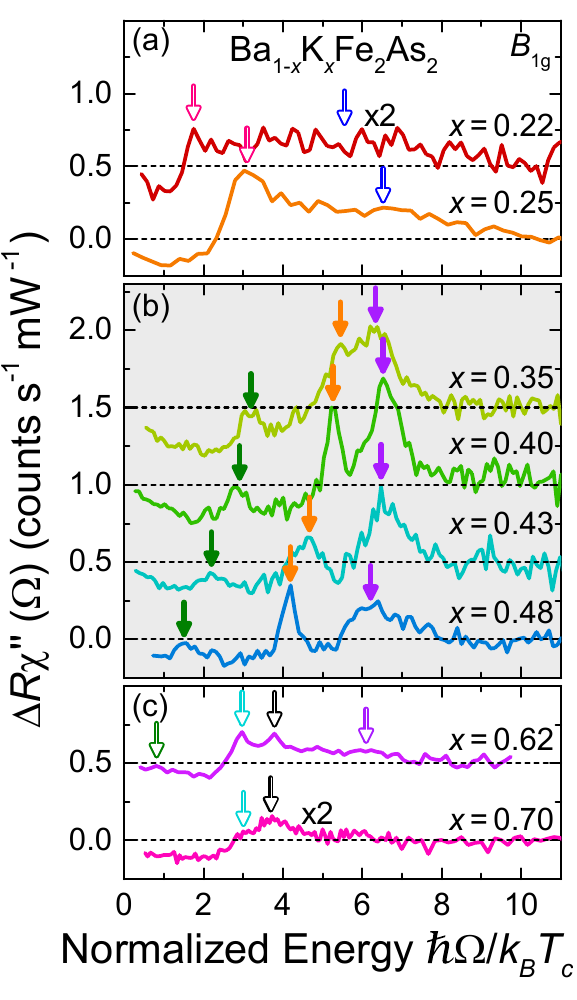}
  \caption{Difference spectra $\Delta R\chi^{\prime\prime}(\Omega,x)$ in $B_{1g}$ symmetry. The energy scale is normalized to the respective $T_c$ values of the differently doped samples. The intensities are off-set, the dashed horizontal lines mark zero. The purple arrows indicate the pair-breaking features at high energy. Orange and green arrows mark the BS modes pulled off of the energy gap. The open black arrow indicates the additional maxima observed at $x=0.62$ and 0.70. Open symbols indicate features with tentative assignment.
  }
  \label{fig:B1g-Tc_all}
\end{figure}

Figure~\ref{fig:B1g-Tc_all} shows the entire set of difference spectra in $B_{1g}$ symmetry. Away from optimal doping the experiments become more difficult to analyze. At $x=0.62$ there are still three superconductivity-induced structures at low temperature similar to those around optimal doping. The low-energy peak is very weak at $x=0.62$ (open green arrow), if existent at all, and is definitely unobservable at $x=0.70$. We believe that it is related to the lower BS mode but find it difficult to furnish experimental proof for this hypothesis.

At $x=0.22$ and 0.25 $\Delta\chi^{\prime\prime}(\Omega,T)$ vanishes only at $10\,k_BT_c$. Consequently, the shoulders developing in the energy range around 100--150\,cm$^{-1}$ ($4-10\,k_BT_c$; open blue arrows in Fig.~2) are likely to have a relationship to the spin density wave (SDW). Then, the prominent maxima at 30 and 65\,cm$^{-1}$ ($3\,k_BT_c$), which show little or no temperature dependence [for $x=0.25$ see supplemental Fig.~\ref{sfig:pol-all}\,(b1--b3)] similar to the pair-breaking peaks studied at optimal doping $x=0.40$ \cite{Bohm:2014}, are probably  gap excitations on the outer hole band which is less affected by the SDW forming below approximately 80\,K. Very speculatively one could also interpret them in terms of fluctuations of the superconducting order parameter (``Higgs modes'') activated by the presence of the SDW \cite{Pekker:2015}. The other bands are expected to participate in the SDW formation as they are better nested and are likely to be gapped out at least partially already above $T_c$.

\section{Symmetry dependence for $0.35 \le x \le 0.48$}
\begin{figure*}[htb]
  \centering
  \vspace{1mm}
  \includegraphics[scale=1]{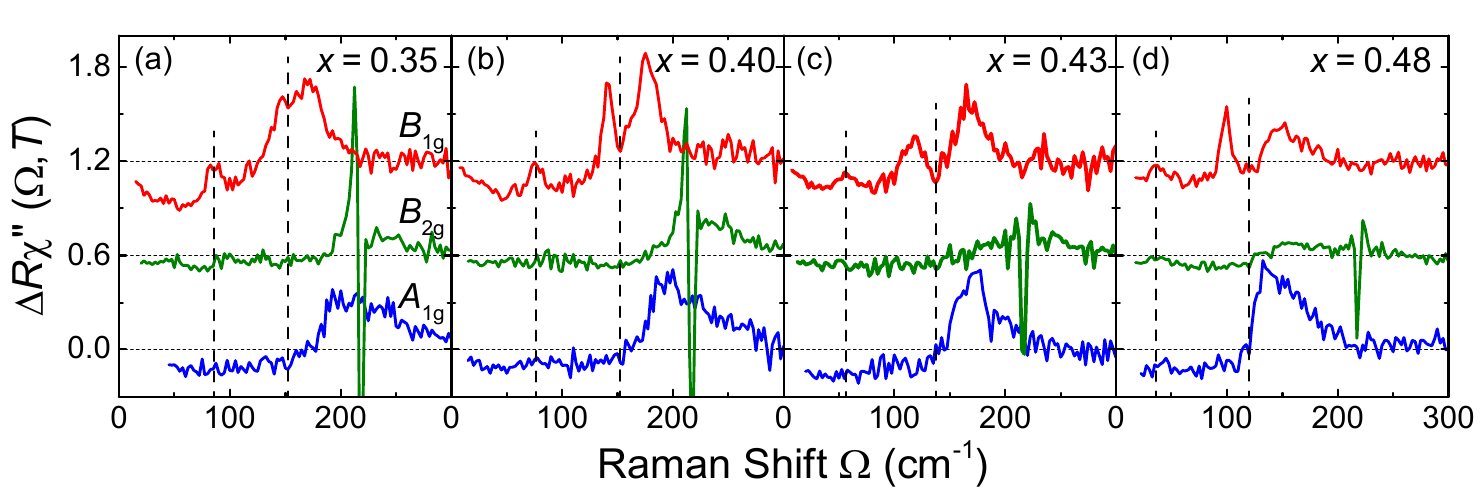}
  \caption{Difference spectra $\mathrm{\Delta R\chi^{\prime\prime}(\Omega,x)}$ for all symmetries close to optimal doping. The dashed vertical lines are intended to demonstrate the uniqueness of the low-energy $B_{1g}$ peak.
  }
  \label{sfig:diff_sym_dop}
\end{figure*}

In Fig.~\ref{sfig:diff_sym_dop} we show the difference spectra for the doping range around $x=0.40$. For all doping levels around optimal doping a pronounced narrow mode below 100\,cm$^{-1}$ exists only in $B_{1g}$ symmetry. In the other symmetries there are only broad shoulders at slightly different energies or no structures at all. For this reason, the mode cannot be a signature of the small gap on the outer hole band which would appear also in the $A_{1g}$ and possibly in the $B_{2g}$ spectra.

\section{Pair breaking and final-state interaction}
In addition to the usual pair-breaking features close to the gap energy $2\Delta$ ${\rm Ba_{1-x}K_{x}Fe_2As_2}$ has two narrow modes inside the gap which originate from the interaction between the electrons of the broken pair as described first by Bardasis and Schrieffer \cite{Bardasis:1961}. For demonstrating the final state interaction to be important the spectra were compared to a phenomenological prediction as described in detail in Ref.~\onlinecite{Bohm:2014}. In brief, the band-dependent gap values are derived from the spectra in $B_{2g}$ and $A_{1g}$ symmetry determining also the bare $B_{1g}$ spectra. Then, the final-state interaction between the electrons of a broken pair is cranked up until the experimental $B_{1g}$ spectra can be reproduced. The closer the coupling $\lambda_d$ in the subdominant $d$-wave channel approaches that in the dominant pairing channel $\lambda_s$ the larger the binding energy $E_b$ of the exciton-like state. Here, $E_b$ means the difference between the gap edge and the position of the collective mode $\Omega_{\rm BS}$. The results of a full 3D phenomenological description are shown in Fig.~\ref{sfig:fits} for the weaker $d(2)$-wave channel BS(2) and doping levels $0.35\le x \le 0.48$.

For the full 3D calculation we use a phenomenological eigenvector $g({\bf k})$ since the fRG and RPA results for $g({\bf k})$ exist only for $k_z=0$. The relative coupling strength $\lambda_d/\lambda_s$ increases monotonically with doping, as plotted in Fig.~3. The values used for the gaps are compiled in Table \ref{tab:gaps}. The phenomenology shows also that the gap on the outer hole band can indeed be observed in all symmetries in the range below 80\,cm$^{-1}$ but the predicted intensity of the pair-breaking maximum in $B_{1g}$ symmetry is at least an order of magnitude too weak to explain the experiment (Fig.~\ref{sfig:fits}).
\begin{table*}[htb]
  \caption{Parameters of the phenomenological fits. We compile the fitting parameters for the phenomenological fits applied to the $B_{1g}$ results of $\rm Ba_{1-x}K_{x}Fe_2As_2$ at doping levels as indicated. All energies are given in units of cm$^{-1}$. According to Ref.~\onlinecite{Bohm:2014} the pockets are labeled as h1 (inner hole pocket), h2 (middle hole pocket), h3 (outer hole pocket), e1 (outer electron pocket), and e2 (inner electron pocket).
  }
  \vspace{1mm}
  \centering
  \begin{tabular}{c c c c c c c c c c c c}
  \hline\hline\\[-2ex]
  $x$& $\Delta\mathrm{_{min}^{h1}}$ & $\Delta\mathrm{_{max}^{h1}}$ & $\Delta\mathrm{_{min}^{h2}}$ & $\Delta\mathrm{_{max}^{h2}}$ & $\Delta\mathrm{_{min}^{h3}}$ & $\Delta\mathrm{_{max}^{h3}}$ & $\Delta\mathrm{_{min}^{e1}}$ & $\Delta\mathrm{_{max}^{e1}}$ & $\Delta\mathrm{_{min}^{e2}}$ & $\Delta\mathrm{_{max}^{e2}}$ & $\Omega_{BS}$\\
  \hline\\[-2ex]
  $0.35$ &180&195&150&256&80&90&165&212&180&210&148\\
  $0.40$ &155&197&170&258&68&80&171&211&177&186&141\\
  $0.43$ &148&162&150&250&50&58&158&210&160&172&119\\
  $0.48$ &123&135&135&150&32&38&138&161&123&133&100\\
  \hline
  \hline
  \end{tabular}
  \label{tab:gaps}
\end{table*}

This discrepancy inspired us to search for an alternative explanation in terms of a second BS mode as described in Ref.~\onlinecite{Maiti:2016}. Since the eigenvectors of the two $d(i)$-wave channels are orthogonal the BS mode for the stronger subleading channel BS(2) can be fitted independently [Fig.~\ref{sfig:fits}\,(e)]. Here the model function is calculated only in 2D since we know the eigenvectors only for $k_z=0$. By using the bare 2D eigenvectors of the subleading channels the fits around $2\Delta$ become worse than in the case and the full analysis. We will address this problem in an upcoming publication.

\begin{figure}[htb]
  \centering
  \vspace{1mm}
  \includegraphics[scale=1]{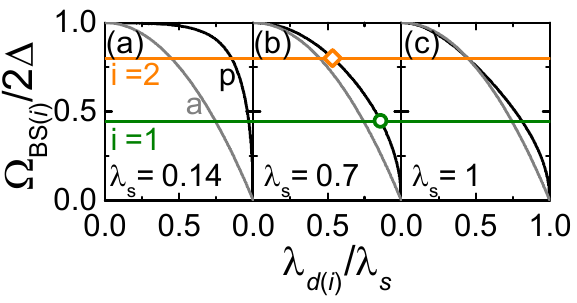}
  \caption{Determination of the relative coupling strengths. The relation between the positions of the BS modes and the ratio of the coupling strength $\lambda_{d(i)}/\lambda_s$ for $x=0.4$ is shown. The horizontal green and  orange lines represent the positions of the BS(1) and BS(2) modes, respectively. The curves show the numerically determined relations between the $E_{b(i)}/2\Delta$ and $\lambda_{d(i)}/\lambda_s$ for $\lambda_s$ as indicated. The limiting case (see text) $\lambda_z = 0$ (black) is shown along with the approximation $\sqrt{E_{b(i)}/2\Delta} = \lambda_{d(i)}/\lambda_s$ (grey). For our analysis we used $\lambda_s=0.7$ and $\lambda_z = 0$ (green circle and orange diamond in (b)).
  }
  \label{sfig:pole}
\end{figure}

The ratio $\lambda_{d(i)}/\lambda_s$ in the fits \cite{Scalapino:2009} and in the single-band results of Monien and Zawadowski \cite{Monien:1990} depends on $\lambda_s$. We use $\lambda_s = 0.7$ as an approximation inspired by the Eliashberg analysis in Refs. \onlinecite{Kuroki:2009a,Ikeda:2010}.
\begin{figure}[htb]
  \centering
  \vspace{1mm}
  \includegraphics[scale=1]{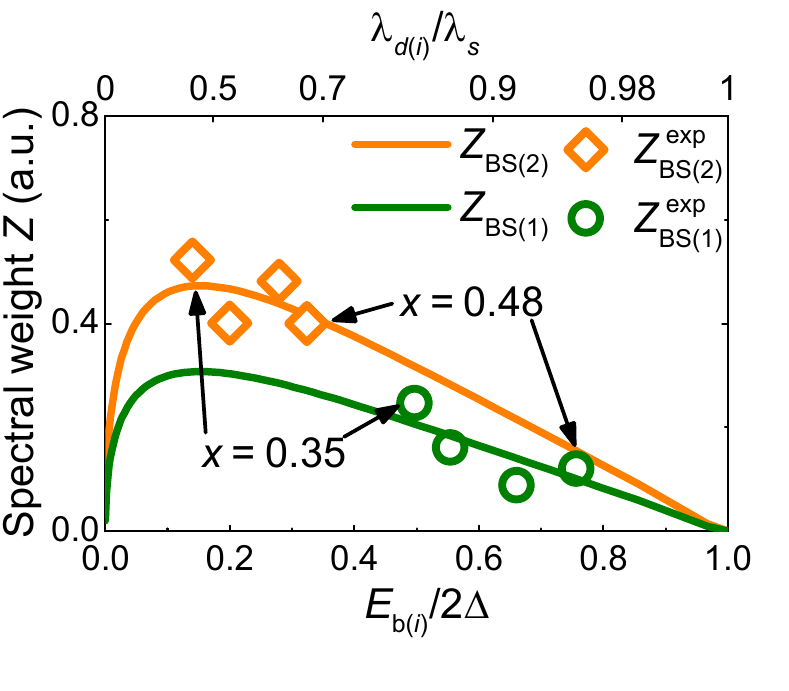}
  \caption{Intensities of the collective modes. The intensities of BS(1) and BS(2) are shown as a function of the relative coupling strength $\lambda_{d(i)}/\lambda_s$ ($i=1,\,2$; $\lambda_s = 0.7$) for the two sub-leading channels). The orange diamonds and green circles represent the experimentally obtained spectral weights of the strong- ($E_{b(1)}\simeq 2\Delta$) and weak-coupling ($E_{b(2)} \ll 2\Delta$) BS modes, respectively. The solid orange and green lines indicate the theoretically expected integrated spectral weight in the BS mode, with $Z_\mathrm{BS(1)}=0.65Z_\mathrm{BS(2)}$.}
  \label{sfig:SW}
\end{figure}

We cannot derive the absolute magnitude of $\lambda_s$ from our experiments. However, we use the experimental positions of the two BS modes and determine $\lambda_{d(1)}$ and $\lambda_{d(2)}$ for three different values of $\lambda_s$ as shown for $x=0.4$ in Fig.~\ref{sfig:pole}. We evaluated the limiting case of $\lambda_z=0$ (black) where $\lambda_z$ is the coupling in the particle-hole channel. For $\lambda_s\simeq 1$, $\sqrt{E_{b}/2\Delta} = \lambda_d/\lambda_s$ (grey) is a useful approximation \cite{Kretzschmar:2013,Bohm:2014} independent of $\lambda_z$. The best agreement between $\lambda_{d(i)}/\lambda_s$ and the ratio derived from fRG and RPA is found for $0.7 \le \lambda_s \le 1$ in agreement with the Eliashberg result \cite{Kuroki:2009a,Ikeda:2010}. For obtaining the numerical value we set $\lambda_z=0$.
\begin{figure}[htb]
  \centering
  \vspace{1mm}
  \includegraphics[scale=1]{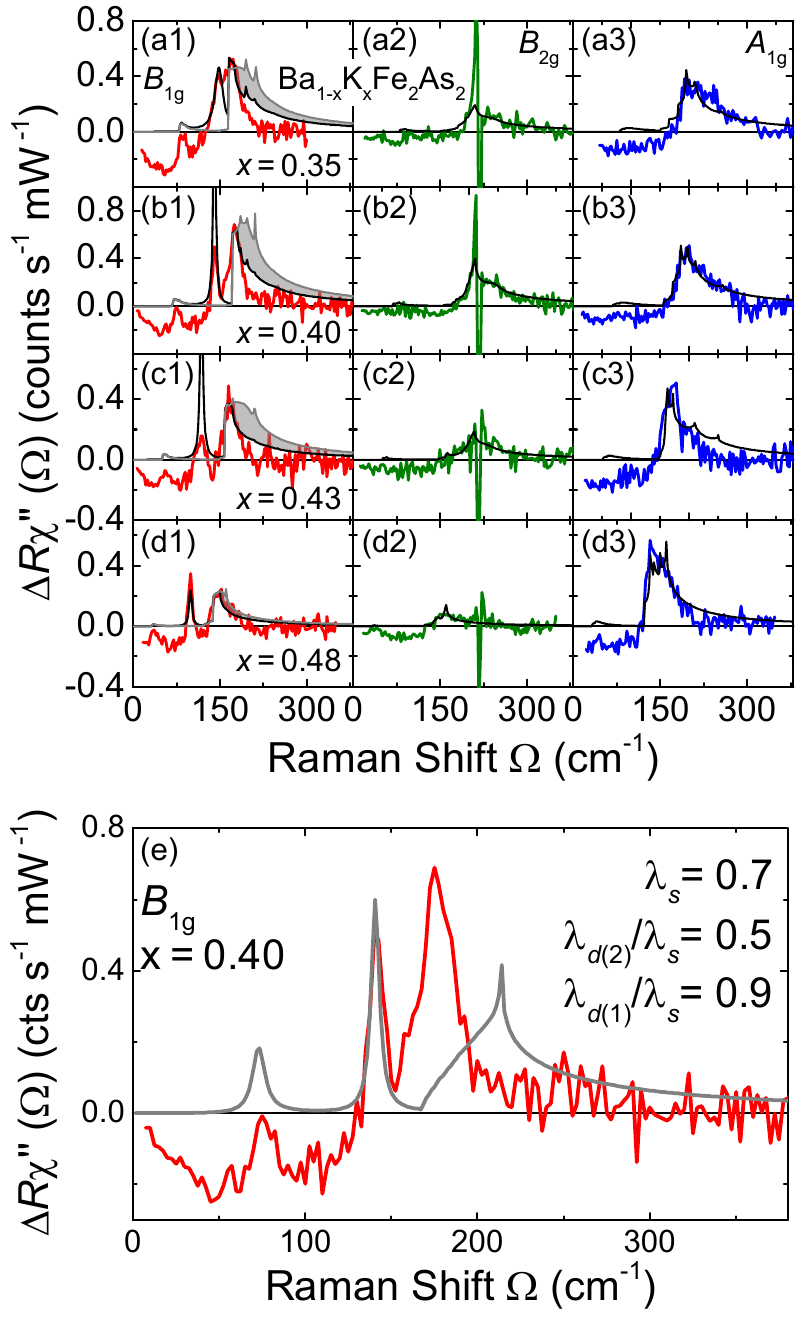}
  \caption{Difference spectra and 3D phenomenological fits. The fits are applied to the data as described in detail in Ref.~\onlinecite{Bohm:2014}. (a1 - d1) The grey-shaded area indicates the spectral weight that is transferred from the pair-breaking peak into the collective mode. (e) 2D model fit with two BS modes. Also here $\lambda_s = 0.7$ is used yielding ratios of $\lambda_{d(1)}/\lambda_s=0.9$ and $\lambda_{d(2)}/\lambda_s=0.5$.
  }
  \label{sfig:fits}
\end{figure}
Na\"{i}vely one would expect the spectral weight in the BS modes $Z_{\rm BS(i)}$ to increase continuously along with $E_{b(i)}/2\Delta$ or, in a different way, with $\lambda_{d(i)}/\lambda_s$. However, $Z_{\rm BS(i)}$ has a maximum already at small $\lambda_{d(i)}/\lambda_s$ as the position of the pole itself depends on the ratio $E_{b(i)}/2\Delta$ or $\lambda_{d(i)}/\lambda_s$ \cite{Monien:1990,Scalapino:2009}. For this reason the BS mode corresponding to the stronger $d(1)$ channel ($E_{b(1)}\simeq 2\Delta$) has a weaker intensity than the other one ($E_{b(2)} \ll 2\Delta$). In Fig.~\ref{sfig:SW} we show $Z_{\rm BS(i)}$ of the two BS modes for $0.35 \le x \le 0.48$. The experimental values for $Z_{\rm BS(1)}$ and $Z_{\rm BS(2)}$ are adjusted to the theory curves using the same factor. The theoretical curves for BS(1) and BS(2) are different because of the different eigenvectors $g_{d(1)}({\bf k})$ and $g_{d(2)}({\bf k})$. Since $g_{d(1)}({\bf k})$ has more nodes the Fermi surface integral is smaller than for $g_{d(2)}({\bf k})$.

\clearpage

\end{appendix}

\end{document}